\documentclass[aip,jcp,reprint]{revtex4-1}
\usepackage{amsmath}
\usepackage{amssymb}
\usepackage{array}
\usepackage{booktabs}
\usepackage{graphicx}
\usepackage[english]{babel}
\usepackage{lipsum}
\usepackage{longtable}
\usepackage{soul}
\usepackage{color}

\newcommand*{\citen}[1]{%
	\begingroup
	\romannumeral-`\x 
	\setcitestyle{numbers}%
	\cite{#1}%
	\endgroup   
}

\begin{document}

\title{ 
Efficient sampling of reversible cross--linking polymers: Self-assembly of single-chain polymeric nanoparticles
}

\author{Bernardo Oyarz\'un}
\affiliation{%
 Universit\'e Libre de Bruxelles (ULB), Interdisciplinary Center for Nonlinear Phenomena and Complex Systems, Campus Plaine, CP 231, Blvd.\ du Triomphe, B-1050 Brussels, Belgium
}
\email{boyarzun@ulb.ac.be, bmognett@ulb.ac.be}
\author{Bortolo Matteo Mognetti }
\affiliation{%
 Universit\'e Libre de Bruxelles (ULB), Interdisciplinary Center for Nonlinear Phenomena and Complex Systems, Campus Plaine, CP 231, Blvd.\ du Triomphe, B-1050 Brussels, Belgium
}

\begin{abstract}
		We present a new simulation technique to study systems of polymers functionalized by reactive sites that bind/unbind forming reversible linkages. Functionalized polymers feature self--assembly and responsive properties that are unmatched by systems lacking selective interactions.
		The scales at which the functional properties of these materials emerge are difficult to model, especially in the reversible regime where such properties result from many binding/unbinding events.
		This difficulty is related to large entropic barriers associated with the formation of intra--molecular loops. 
		In this work we present a simulation scheme that sidesteps configurational costs by dedicated Monte Carlo moves capable of binding/unbinding reactive sites in a single step. 
		Cross-linking reactions are implemented by trial moves that reconstruct chain sections attempting, at the same time, a dimerization reaction between pairs of reactive sites. 
		The model is parametrized by the reaction equilibrium constant of the reactive species free in solution. This quantity can be obtained by means of experiments or atomistic/quantum simulations.
		We use the proposed methodology to study self-assembly of single--chain polymeric nanoparticles, starting from flexible precursors carrying regularly or randomly distributed reactive sites. { During a single run, almost all pairs of reactive monomers interact at least once.}
		We focus on understanding differences in the morphology of chain nanoparticles when linkages are reversible as compared to the well studied case of irreversible reactions. 
		Intriguingly, we find that the size of regularly functionalsized chains, in good solvent conditions, is non--monotonous as a function of the degree of functionalization. 
		We clarify how this result follows from excluded volume interactions and is peculiar of reversible linkages and regular functionalizations.
\end{abstract}

\maketitle

\section{Introduction}

One of the challenges of nanophysics is the design of functional materials capable, for instance, to feature different collective behaviors or to respond in desired ways when triggered by specific external conditions. In bottom-up approaches such designs are based on the possibility of controlling interactions between molecular groups. The bottom-up program has been boosted by the use of materials, like polymers or colloids, conjugated with reactive complexes capable of forming supramolecular linkages \cite{Lehn1988}. In these systems the interaction between polymers and colloids is controlled by the degree of functionalization or the affinity between reactive complexes. Supramolecular interactions are currently used, for instance, to program state-dependent interactions in systems of functionalized colloids \cite{Jones2015}, or to design selective vectors for drug delivery applications \cite{Kiessling2000}. Supramolecular interactions are also found in many biological systems. For instance, ligand-receptor interactions control inter-membrane adhesion and initiate signaling cascades, while the mesoscopic structure of chromatin in eukaryotic cells is regulated by proteins that cross-link the genetic fiber \cite{BookMolecularBiology}. 

Optimization of control (e.g.~thermodynamic) parameters in bottom-up systems necessitates computational platforms to predict collective properties. However, modeling the functional behavior of supramolecular systems is difficult because it requires merging atomistic descriptions, needed to properly describe reacting complexes, with large-scale simulations. In general, reactions between functionalized polymers are hampered by configurational costs due to the fact that backbones carrying complementary moieties have to be close enough to let the complexes react. These entropic terms, that in typical polymeric networks can be of the order of few tens of $k_B T$ (e.g.\ \cite{Marenduzzo2006,melting-theory1}), { result} in unaffordable simulation times when attempting, for instance, to study polymer cyclization by using algorithms based on physical dynamics. In systems of functionalized particles, theoretical and simulation methods have been developed to calculate effective particle interactions \cite{melting-theory1,DNACCtheoryreview} starting from hybridization free energies of tethered polymers tipped by reactive complexes. Effective interactions can then be used to simulate many particle properties like self-assembly. However, the use of effective interactions is not feasible when studying conjugated polymers in view of the many possible configurations taken by polymer backbones, and the significant effect that supramolecular linkages have on the actual morphology of the chain. Note that, as already highlighted when calculating secondary structures of nucleic acid strands \cite{Robert2007}, analytical treatments of entropic configurational terms in chains that are cross-linked multiple times (see Fig.\ \ref{Fig:Figure1} (a)) are not available even when the sampling is limited to ideal and unpseudoknotten configurations \cite{Robert2007}.

In this paper we present a general algorithm to simulate polymer networks forming reversible supramolecular linkages (e.g.\ \cite{Zhong2016,Wang2017,Aida2012,Mohan2010,Biffi2013,hugouvieux2016,stuart2010,megariotis2016,schmid2013,Jacobs2016}).
These are complexes that are formed, {\em via} hydrogen bonding \cite{Sijbesma1997,Cordier2008,Tang2008}, metal-ligand interactions \cite{Beck2003}, hydrophobic interactions \cite{Rauwald2008}, $\pi$--stacking \cite{Hoeben2005}, {\em etc}\cite{Lehn1995}. The scope of the proposed method is to efficiently sample between network topologies corresponding to different pairs of reacted complexes. Ultimately, this will permit to access the lengthscales at which these materials function and, in particular, to study how they self-assemble. Kinetic bottlenecks, limiting the number of binding/unbinding events between complexes, are overcome by dedicated Monte Carlo moves in which complexes tethered to backbones are bound/unbound in a single step of the algorithm. The methodology is based on configurational bias moves~\cite{Siepmann1992,Mooij1992,dePablo1992,Mooij1994} powered by topological jumps as previously tested when studying hybridization of tethered single-stranded DNA carrying reactive sticky ends, and adsorption of ideal chains on functionalized interfaces \cite{DeGernier2014}. We stress that the proposed methodology is general allowing the study of different types of supramolecular linkages (as specified by size, strength, or directionality) and how they influence large-scale properties of the aggregates. { However, non--local Monte Carlo moves cannot reproduce the dynamics of the system and can break topological constraints  \cite{Padding2001,Tzoumanekas2006,Micheletti2011,Ramirez2017}.  On the other hand, the non-crossability of bond segments can be enforced more efficiently when using Molecular Dynamics simulations. 
}

To test the method, we study functionalized single chains forming intra-molecular reversible linkages. These systems are being used to self-assemble single-chain polymeric nanoparticles (SCPNs)\cite{Altintas2012,Altintas2016,Moreno2017} with applications in Materials Science (e.g.\ sensors for metal ions\cite{Martijn2012} or catalysts \cite{Neumann2015}) and Nanomedicine  (e.g.\ presbyopia treatment \cite{Liang2017}). Existing numerical works on folding of SCPNs rely on Molecular Dynamics simulations of atomistic \cite{Liu2008,Mondello1994,Ferrante2012} or coarse-grained models\cite{Moreno2017,Moreno2013,LoVerso2014,LoVerso2015,Bae2017,Englebienne2012} in which interactions between functional groups are modeled by means of bead-and-spring interactions. With the exception of theoretical studies \cite{Pomposo2017} most of simulation works take the irreversible limit, in which once two functional groups are found close to each other a permanent intra-molecular linkage is added between them \cite{Moreno2013,LoVerso2014,LoVerso2015,Bae2017}. Instead, in this work we study the emergent structure of SCPNs as obtained starting from a precursor in good solvent conditions and by letting the chain to explore different connectivity states characterized by different sets of pairs of linked complexes, see Fig.~\ref{Fig:Figure1} (a). With respect to existing methods, we find that our algorithm can explore a large number of connectivity states in a single run. This results in averaged connectivity maps (see Fig.\ \ref{Fig:Figure1} (b) for the example of a single and Fig.\ \ref{Fig:Figure9} for an averaged connectivity map) that resemble what found in high--throughput experiments on chromatin \cite{Lamb2006}.
\\
Most of the existing literature has studied the morphology of SCPNs as a function of systems' design parameters (degree of functionalization and molecular weight of the polymer\cite{Moreno2013,LoVerso2014,terHuurne2017,Stals2014,Pomposo2014}) using different experimental settings (employing linkers \cite{Perez-Baena2014}, crowders \cite{Formanek2017}, or selective solvents \cite{LoVerso2015}).
Starting from precursors in good solvent conditions, it has been highlighted how excluded-volume effects favor short loops hampering chain compaction \cite{Moreno2016}. 
In this paper we confirm this result in the case of reversible linkages. 

We then compare the size of SCPNs as obtained using regularly and randomly functionalized precursors.  We find that in the second case the size of the SCPNs is re--entrant with respect to the degree of functionalization. Interestingly, this re--entrant behavior disappears in the irreversible limit where linkages once formed are quenched and the chains cannot minimize excluded volume effects.

Here is the plan of the work. In Sec.~\ref{Sec:Model} we introduce the model. In Secs.~\ref{Sec:Algorithms} and \ref{sec:alg-val} we describe and validate, respectively, the simulation algorithm. Further details about the method are reported in Appendix~\ref{App:reac} and~\ref{App:DetBal}, while in Appendix~\ref{App:Analytic} we report analytic calculations that have been used to validate the program. In Sec.~\ref{Sec:Results} we present our results about self-assembly of SCPNs obtained starting from homofunctional precursors. Finally in Sec.~\ref{Sec:Conc} we summarize our results and itemize future directions of research. 

\section{Methodology}\label{Sec:Model}

 \begin{figure}
	\centering
	\includegraphics[width=0.5\textwidth]{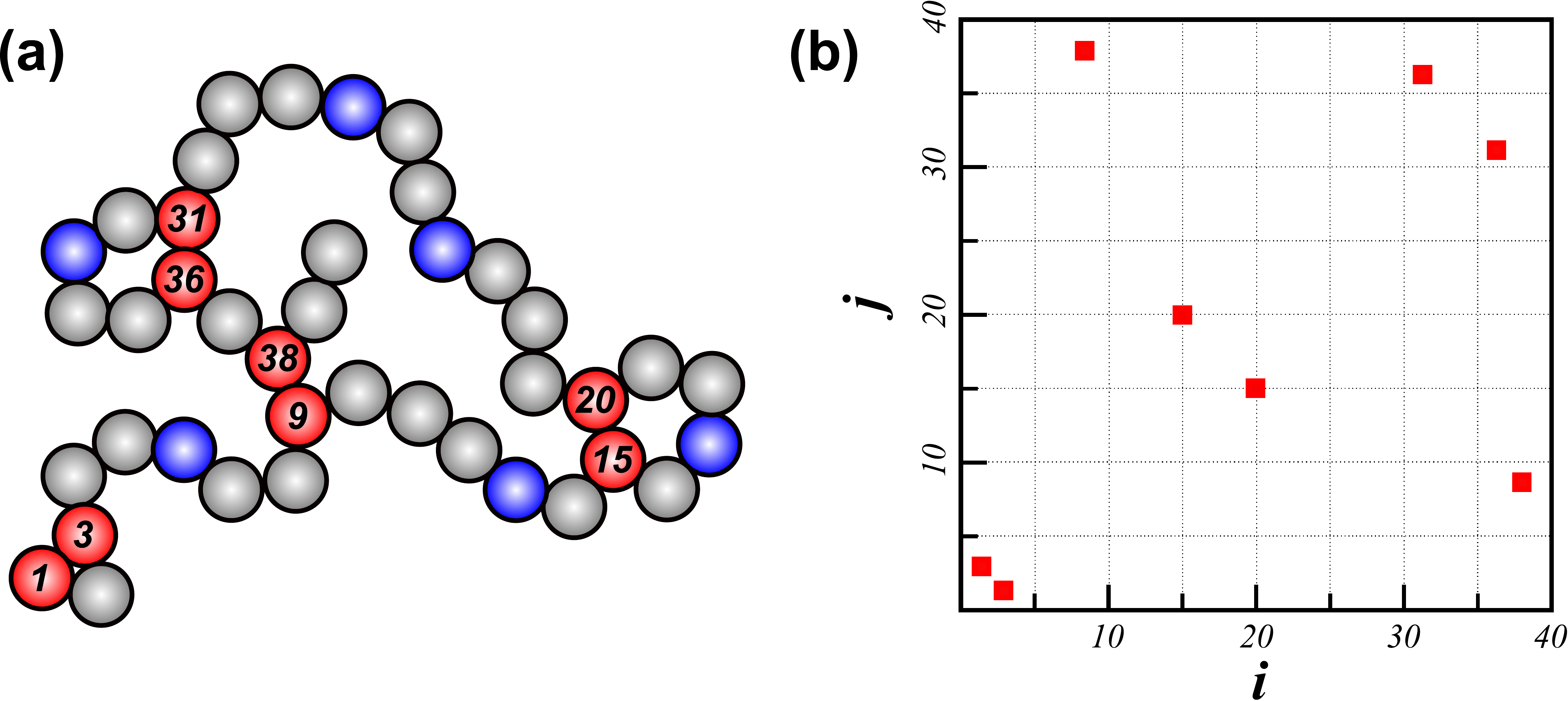}
	\caption{(a) Schematic representation of a cross-linked single-chain polymeric nanoparticle. Blue and red monomers depict free and cross-linked reactive sites, where the numbers highlight the identity of the cross-linked monomers. (b) Connectivity map $\nu$ listing the pairs of cross-linked monomers for the polymer shown in (a).
	}
	\label{Fig:Figure1}
\end{figure}

In this work, we propose a general method to study the morphology of cross-linking polymer networks. Modeling the atomistic details of polymers, and their reactive sites, is not affordable when studying large-scale morphology of functional polymers. For this reason, chemical details of the polymer backbone, reactive sites, and bound complexes are treated at the coarse--grained level. In our study, polymer precursors are represented as chains with fixed bond length (equal to $\sigma$) made of $N_C$ monomers, $N_R$ of which ($N_R \le N_C$) are reactive and can form cross-linking complexes following a dimerization reaction,
 
\begin{eqnarray}
2A \rightleftharpoons A_2 \, \, .
\label{Eq:Rx}
\end{eqnarray}

\noindent
The type of chemical reaction determines the connectivity states featured by the system. Although the model derivation presented below is based on Eq.~\ref{Eq:Rx}, other types of reactions can be considered using this method.
 
Following previous strategies in modeling DNA functionalized colloids \cite{DNACCtheoryreview,melting-theory1}, chemical details of reacting groups are parametrized by the thermodynamic properties of diluted free reacting species (for the reaction of Eq.\ \ref{Eq:Rx} a gas of $A$ monomers and $A_2$ dimers at equilibrium). The partition function of an $A$ and $A_2$ molecule free in solution is defined as $V q_A$ and $V q_{A_2}/2 $, respectively. $q_A$ and $q_{A_2}$ lump all momentum and internal degrees of freedom contributions of the corresponding molecules moving in a volume $V$. These terms are directly linked to the equilibrium constant $K_{\textrm{eq}}$ of the dimerization reaction by $K_{\textrm{eq}}=(q_{A_2}/2)/(q_A)^2$ (see Appendix \ref{App:reac} for a derivation). This relation allows the parametrization of the coarse-grained model using quantities that are readily accessible by experiments or quantum mechanical calculations.

The aim of the method proposed here is to efficiently sample reversible cross-linking states of polymer networks. A cross-linked microstate is characterized by a connectivity matrix $\nu$ listing all pairs of reacted complexes, Fig.~\ref{Fig:Figure1}~(b). A given connectivity matrix is { translated} into a set of distance constraints between all pairs of reacted monomers. For a given precursor, the partition function describing the properties of a cross-linking system is then written as,
\begin{eqnarray}
\label{Eq:Partfunc}
Z &=&  \sum_{\nu} Z_{\nu} = \sum_\nu \left(  q_{A_2} \over \Omega_0 \right)^{n_{\nu}} \; q_A^{(N_R- 2 n_{\nu})}  {\cal Z}_\nu
\\
{\cal Z}_\nu &=&  \int_{\varphi} \mathrm{d} {\bf r}  \exp[-\beta \, {\cal U} ({\bf r}) ] \prod_{i<j | m_{ij}\neq 0} \delta (|{\bf r}_i - {\bf r}_j| - {\sigma}),
\nonumber
\end{eqnarray}

\noindent
where the sum is taken over all possible connectivity matrices $\nu=\{m_{ij}\}$. In particular $m_{ij}=1$  ($m_{ij}=0$) if reactive monomers $i$ and $j$ are (not) linked, and $m_{ij}=0$ if $i$ or $j$ are not reactive. $n_\nu$ is the number of formed linked complexes for a given $\nu$, $n_\nu=\sum_{ij} m_{ij}/2$. In Eq.\ \ref{Eq:Partfunc} the delta functions enforce the fixed distance constraints between each couple of reacted dimers listed by a given connectivity matrix. $\Omega_0$ is the angular contribution to the partition function of a reacted complex ($\Omega_0=4\pi\sigma^2$), which is included in Eq.~\ref{Eq:Partfunc} to compensate for the fact that the rotational degrees of freedom of $A_2$ dimers are included twice, in the definition of ${\cal Z}_\nu$ and in $q_{A_2}$ (see Appendix~\ref{App:reac}). 
In Eq.\ \ref{Eq:Partfunc}, ${\cal U}({\bf r})$ accounts for monomer-monomer interactions, while $\varphi$ refers to the space of monomer coordinates restricted to the fixed bond constraint $| {\bf r}_{i+1} - {\bf r}_i|=\sigma$. As an example in Appendix \ref{App:Analytic} we explicitly calculate the partition function $Z$ of an ideal chains (${\cal U}=0$) functionalized by four reactive sites ($n_\nu=0,\, 1, \, 2$).

To sample Eq.\ \ref{Eq:Partfunc} we need to devise efficient Monte Carlo methods capable of sampling between different connectivity microstates. This is done in the next section.

\section{ Algorithms }
\label{Sec:Algorithms}

In this section we present MC algorithms capable of sampling between chain configurations featuring different connectivity matrices of the model defined by Eqs.~\ref{Eq:Rx} and~\ref{Eq:Partfunc}, and represented in Fig.~\ref{Fig:Figure1}. These algorithms require the reconfiguration of sections of the polymer to allow the cross-linking reaction to take place. In Sec.~\ref{Sec:Grow} we describe a method to generate polymer sections with fixed end--points based on a Markowian process. Using the method proposed in Sec.~\ref{Sec:Grow}, we develop in Sec.~\ref{Sec:Link} an algorithm that changes the number of cross-linking complexes from $n_\nu$ to $n_\nu+1$ or $n_\nu-1$ (see Eq.\ \ref{Eq:Partfunc}), respectively. Sec.~\ref{Sec:Relax} describes other MC moves that have been used to relax the polymeric network at fixed $\nu$.

\subsection{ Growth of an internal section of the polymer }
\label{Sec:Grow}

In this section, we describe the growing procedure used to reconfigure internal sections of polymer chains. The procedure is based on the growing scheme described in the topological configurational bias Monte Carlo method~\cite{DeGernier2014}, where new chain configurations are grown segment-by-segment between two fixed points following a Rosenbluth scheme~\cite{Rosenbluth1955} guided by the end-to-end probability distribution function of ideal chains. This scheme is similar in spirit to other fix-end biased growing methods used to reconfigure internal sections of a chain~\cite{Dijkstra1994,Pant1995,Escobedo1995b,Vendruscolo1997,Wick2000,Uhlherr2000,Chen2000b,Sepehri2017a}. The proposed method is simple and efficient, avoiding numerical~\cite{Vendruscolo1997} or iterative estimation~\cite{Wick2000,Chen2000b} of the guiding functions.

\begin{figure}
	\centering
	\includegraphics[width=0.45\textwidth]{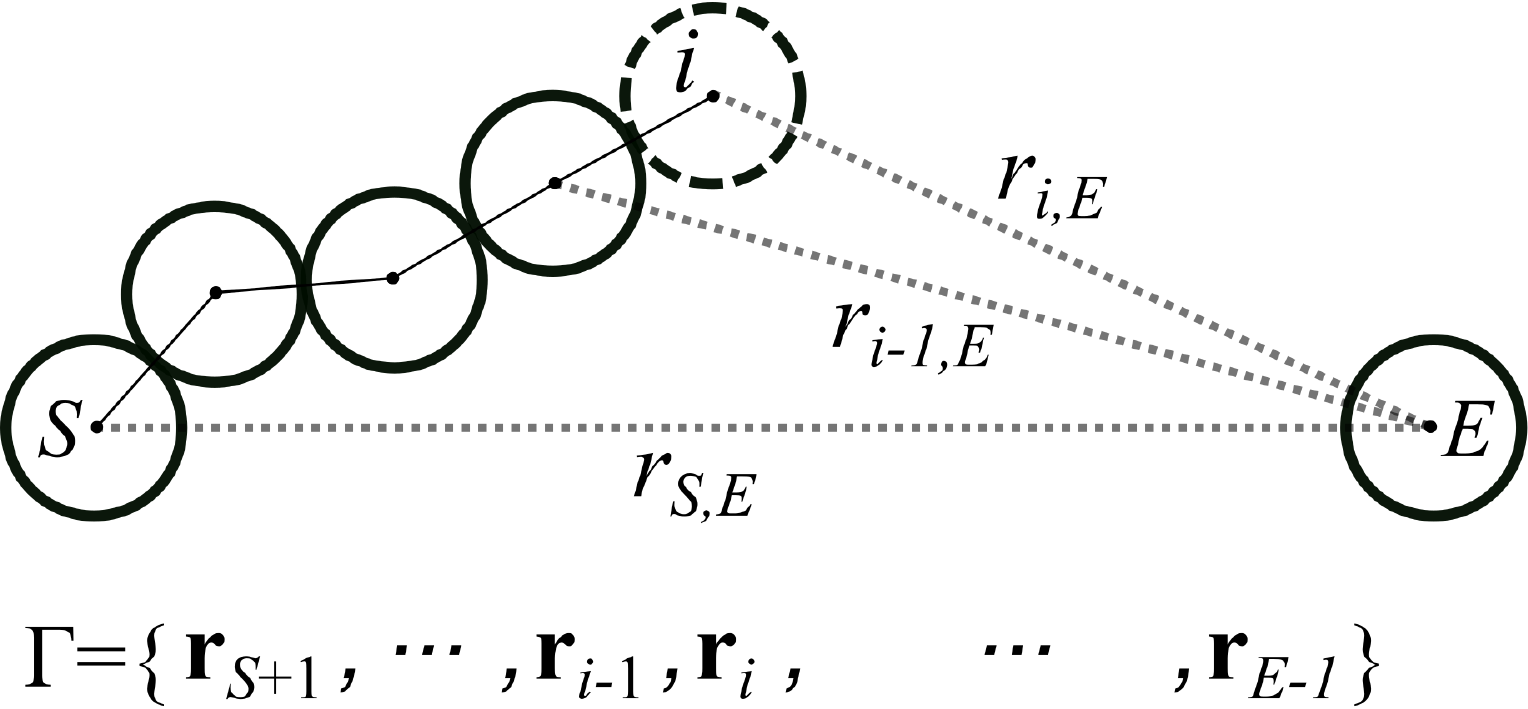}
	\caption{Growing procedure of a chain section $\Gamma$ between two monomers with fixed positions ${\bf r}_S$ and ${\bf r}_E$. The chain section $\Gamma$ is made of $L_{E,S}=E-S$ segments and $N_{E,S}=L_{E,S}-1$ monomers. The chain section is grown segment-by-segment selecting the position of monomer $i$ from a series of trial positions around ${\bf r}_{i-1}$. 
	This selection is biased by the probability densities of ideal chains made of $E-i$ segments with end-to-end distances equal to ${\bf r}_{i,E}$.
	}
	\label{Fig:Figure2}
\end{figure}

In Fig.~\ref{Fig:Figure2} we consider the growth of a chain section ${ \Gamma}$ between monomer $S$ and monomer $E$, whose positions, ${\bf r}_S$ and ${\bf r}_E$, are kept fixed. The indexes $S$ and $E$ identify the monomer number in increasing order, therefore the number of segments and monomers of section $\Gamma$ are  $L_{S,E}=E-S$ and $N_\Gamma=L_{S,E}-1$, respectively.  As previously done for flexible chains~\cite{DeGernier2014}, we use as guiding functions the probability distribution of the end-to-end distance, $r$, of freely--jointed chains made of $L$ segments, $p(r,L)$. An analytic expression of these functions are reported in Refs.~\citen{Treloar1946} and \citen{Yamakawa1971}. The usual configurational bias Monte Carlo (CBMC) algorithm~\cite{Mooij1992,dePablo1992,Siepmann1992} is used in the case that $\Gamma$ comprises one of the two unconstrained ends of the polymer.

The growth proceeds one monomer at a time by the following scheme:\\ 
\noindent
i) At every growth step $i$ ($S<i<E$),  $k$ trial vectors of length $\sigma$ (${\bf u}^{(n)}$, $n=1 ,\cdots, k$) are randomly generated on a sphere centered on the position of the previously grown monomer ${\bf r}_{i-1}$, defining $k$ trial positions of monomer $i$, ${\bf r}_i^{(n)} = {\bf r}_{i-1} + {\bf u}^{(n)}$ with probability,

\begin{eqnarray}
\label{Eq:Gen_k}
\nonumber
\mathrm{Pr}({\bf r}_i^{(n)}) &=& { p(|{\bf r}_E - ({\bf r}_{i-1} + {\bf u}^{(n)}) |, L_{i,E})  \over \int_{\Omega_0} d{\bf u} \; p(|{\bf r}_E - {\bf r}_{i-1} + {\bf u}|,L_{i,E})}\\
&=& { p(r_{i,E}^{(n)}, L_{i,E}) \over \Omega_0 \;   p(r_{i-1,E}, L_{i-1,E}) },
\end{eqnarray}

\noindent
 where $r_{i,E}^{(n)}$ is the end-to-end distance between trial $n$ and the end-point $E$, and $L_{i,E}=E-i$ is the number of segments to reach the end-point $E$ from monomer $i$ (similar definitions hold for $r_{i-1,E}$ and $L_{i-1,E}$).
The second equality in Eq.~\ref{Eq:Gen_k} is peculiar to the case of end-to-end distributions of fully--flexible ideal chains. Practically, we sample Eq.\ \ref{Eq:Gen_k} using a rejection algorithm. Trial positions with an end-to-end distance $r_{i,E}$ larger than $L_{i,E} \cdot \sigma$ are rejected.\\

\noindent
ii) The position of monomer $i$ (${\bf r}_i$) is randomly chosen within the $k$ trials

\begin{eqnarray}
\label{Eq:Sel_k}
\mathrm{Pr}({\bf r}_i) &=& { \exp[-\beta {\cal U}({\bf r}_i ; {\Gamma}_i)] \over W_i } \\
W_i &=& {\sum_{n=1}^k \exp[-\beta {\cal U}({\bf r}^{(n)}_i ; {\Gamma}_i)]},
\nonumber 
\end{eqnarray}

\noindent
where ${\cal U}({\bf r}_i;{\Gamma}_i)$ is the interaction energy of monomer $i$ with ${\Gamma}_i=\{{\bf r}_{S+1},\cdots,{\bf r}_{i-1}\}$, and with monomers not contained in ${\Gamma}$. $W_i$ is the Rosenbluth factor of monomer $i$.\\

\noindent
iii) The previous procedure is repeated until all $N_{\Gamma}$ monomers are placed. Using Eq.~\ref{Eq:Gen_k} and \ref{Eq:Sel_k} the probability of generating a chain section ${\Gamma}=\{ {\bf r}_{S+1}, \cdots, {\bf r}_{E-1} \}$ is given by,

\begin{eqnarray}
\label{Eq:Grow}
\nonumber
\mathcal{P}^{\textrm{gen}}_{\Gamma} &=& {1 \over \Omega_0^{N_{\Gamma}}}\prod_{i=S+1}^{E-1} 
{p(r_{i,E},L_{i,E}) \over p(r_{i-1,E},L_{i,E-1})} \frac{\exp[-\beta \; {\cal U}({\bf r}_i ; {\Gamma}_i)]}{W_i}\\
&=& {1\over \Omega_0^{N_{\Gamma}}} { p(\sigma,1) \over p( r_{S,E} ,L_{S,E})} \frac{\exp[-\beta \; {\cal U}({\Gamma})]}{W_{\Gamma}},
\end{eqnarray}

\noindent
where $r_{S,E}$ is the distance between monomer $S$ and $E$, and $p(\sigma,1)$ is the probability of closing the chain section $\Gamma$, i.e the probability of placing the last monomer at a distance $\sigma$ from $E$. This last probability is simply equal to $1/\Omega_0$ when growing a covalent bond given that we sample chains restricted by the measure $\varphi$ (see Eq.~\ref{Eq:Partfunc}), and equal to $\delta(|{\bf r}_{\alpha} - {\bf r}_{\beta}|-\sigma)/\Omega_0$ in the case that two reactive monomers $\alpha$ and $\beta$ are reversibly cross-linked (see next section).  ${\cal U}({ \Gamma})$ is the interaction energy of the grown chain section with itself and with the remaining of the chain. The Rosenbluth factor of the whole growing process is given by $W_{\Gamma}=\prod_{i=S+1}^{E-1}W_i$. Note that the Jacobian term used in previous works~\cite{Dodd1993,Escobedo1995b,Wick2000} corresponds to $p(r_{E-2,E},2)$. 
The latter term does not appear in ${\cal P}^{\textrm{gen}}_{\Gamma}$ because of the chain simplification in Eq.~\ref{Eq:Grow}.

\subsection{ Binding/unbinding }
\label{Sec:Link}

\begin{figure}
	\centering
	\includegraphics[width=0.5\textwidth]{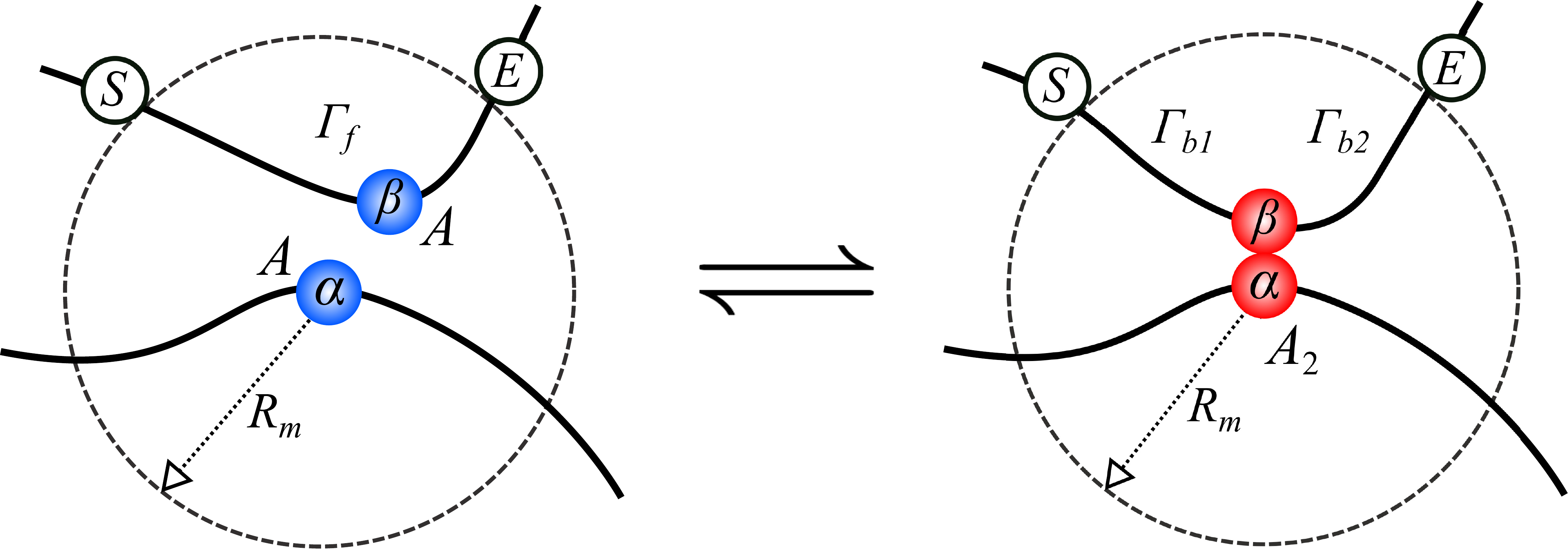}
	\caption{Cross-linking reaction between two reactive sites $A$ to form a cross-linking complex $A_2$. The reactive monomer $\alpha$ defines the center of a sphere $V_m$ of radius $R_m$ where a second reactive monomer $\beta$ is found. The chain section $\Gamma$, which is limited by the monomers $S$ and $E$, is reconfigured from free $\Gamma_f$ to cross-linked $\Gamma_b$ states and backwards.	
	}
	\label{Fig:Figure3}
\end{figure}

\begin{figure}
	\centering
	\includegraphics[width=0.5\textwidth]{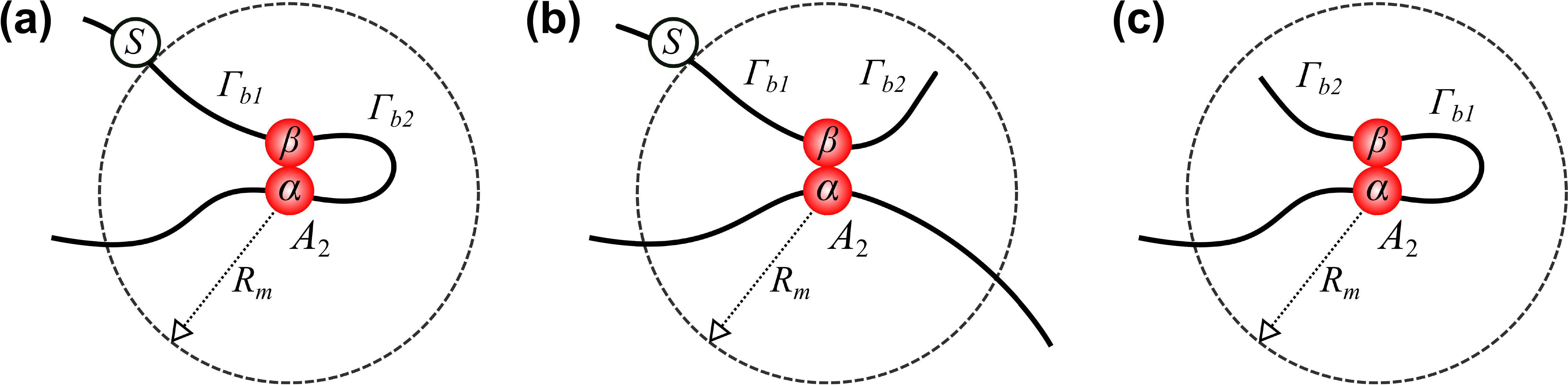}
	\caption{Different growing scenarios starting from a cross-linked configuration: (a) The end monomer (E in Fig.\ \ref{Fig:Figure3}) is equal to one of the reactive monomers. (b) The end monomer is a polymer end. (c) The end monomer is a polymer end while the start monomer (S) is one of the reactive monomers.
	}
	\label{Fig:Figure4}
\end{figure}

The connectivity of polymeric networks is changed by a Monte Carlo algorithm that attempts to bind two reactive monomers $A$ or to unbind one cross-linked complex $A_2$.
Forming a cross-linking complex requires the re-arrangement of the polymer backbone bringing two reactive sites to a distance equal to one bond length $\sigma$. In our method, we attempt simultaneously to rearrange the configuration of a chain section while attempting a cross-linking reaction. Fig.~\ref{Fig:Figure3} shows schematically the procedure, where one reactive site (identified by $\alpha$) is kept fixed while the configuration of a chain section $\Gamma$ containing a second reactive site $\beta$ is changed. In particular, $\Gamma$ is constrained to lay inside a sphere $V_m$ of radius $R_m$ centered on $\alpha$, limited by monomers $S$ and $E$ that
are found just outside $V_m$. The introduction of the sphere $V_m$ is necessary to define a proximity space where the reaction can take place. The limiting monomers $S$ and $E$ are determined starting from $\alpha$ and moving along the chain backbone in the direction of $\beta$. Note that $\alpha$ can coincide with $S$ or $E$, and that there is only one limiting monomer in the case that $\Gamma$ includes one of the two ends of the polymer. In this last case $\Gamma$ is said to be a dangle terminal. All possible cases are reported in Fig.~\ref{Fig:Figure4}.  It is important to stress that the size of $V_m$ determines solely the efficiency of the algorithm, as shown in Sec.\ \ref{sec:alg-val} and \ref{Sec:Results} (see Fig.\ \ref{Fig:Figure5} and \ref{Fig:Figure10}). In particular, results are unaffected by the particular choice of the radius $R_m$.\\ 

In the case that a linking complex is attempted to form, the chain section $\Gamma$ is reconfigured to a new cross-linked configuration ${ \Gamma}_{b}=\{{\bf r}_{S+1},\cdots,{\bf r}_{\beta},\cdots,{\bf r}_{E-1}\}$ by using a two-step procedure. First, the subchain ${\Gamma}_{b1}=\{{\bf r}_{S+1},\cdots,{\bf r}_{\beta}\}$ is grown using ${\bf r}_S$ and ${\bf r}_{\alpha}$ as starting and end fixed coordinates, placed at a relative distance $r_{S,\alpha}$. We use the algorithm described in Sec.\ \ref{Sec:Grow} in which $E$ is replaced by $\alpha$, and the chain section generated is made of $L_{{S,\beta}}+1$ segments. The extra segment is a consequence of the new bond that is introduced by the cross-linking reaction. After $\Gamma_{b1}$ is grown, the position of the reactive monomer $\beta$ is fixed at a relative distance $r_{\beta,E}$ from monomer $E$. To complete the linking attempt a second subchain ${\Gamma}_{b2}=\{{\bf r}_{\beta+1}, \cdots {\bf r}_{E-1} \}$ made of $L_{\beta,E}=E - \beta$ segments is grown with ${\bf r}_\beta$ and ${\bf r}_E$ as starting and end fixed coordinates (see Fig.\ \ref{Fig:Figure2}).
Adapting Eq.\ \ref{Eq:Grow} to the present two-step linking process, a cross-linked configuration is generated with probability, 

\begin{eqnarray}
\nonumber
\mathcal{P}^{\textrm{gen}}_{\Gamma_b} &=& \mathcal{P}^{\textrm{gen}} ( { \Gamma}_{b1}) \; \mathcal{P}^{\textrm{gen}}( {\Gamma}_{b2} | {\Gamma}_{b1} )\\
\nonumber
&=&{\chi_{V_m}({\Gamma}_b)  \over \Omega_0^{N_{\Gamma}+ 2}} 
\frac{ \delta (  r_{\alpha,\beta} - \sigma)}{p(r_{S,\alpha},{L_{{S,\beta}}}+1)}
\frac{1}{p(r_{\beta , E},L_{{\beta,E}})}
\\
&& \times \; \frac{\exp[-\beta \; {\cal U}({\Gamma}_b)]}{W_{{\Gamma}_b}},
\label{Eq:Gammab}
\end{eqnarray}

\noindent
where $\chi_{V_m}({\Gamma}_b)$ is a test function that restricts $\Gamma_b$ to lay inside $V_m$. $\chi_{V_m}{=}1$ if ${\bf r}_i {\in} V_m; \forall {\bf r}_i {\in} {\Gamma}$, and $\chi_{V_m}{=}0$ otherwise. This constraint is important to guarantee the reversibility of the algorithm that, in particular, requires that the positions and identity of $S$, $\alpha$, and $E$ do not change after a binding/unbinding event. Practically, we treat $\chi$ as an excluded volume interaction when selecting new segments in Eq.~\ref{Eq:Sel_k}.
The interaction energy ${\cal U}({\Gamma}_b)$ is defined as in Sec.\ \ref{Sec:Grow} and takes into account the potential energy between monomers in the chain section $\Gamma_b$ and with the remaining of the chain. We neglect interactions between reversibly cross--linked monomers because such terms already enter in the definition of $q_{A_2}$ (Eq.\ \ref{Eq:Partfunc}) and the equilibrium constant $K_\mathrm{eq}$. The Rosenbluth factor is given by $W_{\Gamma_b}=W_{\Gamma_{b1}}W_{\Gamma_{b2}}$ for the whole growing process where interactions between $\Gamma_{b1}$ and ${\Gamma_{b2}}$ are considered when calculating $W_{\Gamma_{b2}}$. 
In Eq.\ \ref{Eq:Gammab} we have used Eq.\ \ref{Eq:Grow} after setting $p(\sigma,1)=\delta(r_{\alpha,\beta} - \sigma)/\Omega_0$ when growing $\Gamma_{b1}$, as explained in Sec.~\ref{Sec:Grow}.

The scheme of Eq.~\ref{Eq:Gammab} is not symmetrical when exchanging $S$ with $E$ (direction of growth) due to the fact that the growth of $\Gamma_{b2}$ depends on $\Gamma_{b1}$. In particular, the position of the reacting complex $\beta$ is solely determined by the growth of $\Gamma_{b1}$. Such asymmetry is compensated by the acceptance rules as explained below. Alternatively, the position of monomer $\beta$ could be sampled first, followed by a two chain growth from $\beta$ towards $S$ and $E$. This algorithm will be studied in a future contribution. 

When attempting to unbind a cross-linked complex ($A_2$ in Fig.\ \ref{Fig:Figure2}) a new free chain configuration  ${\Gamma}_{f}=\{{\bf r}_{S+1},\cdots,{\bf r}_{E-1}\}$ is grown inside $V_m$ between the starting $S$ and end $E$ points. The probability of generating such a configuration is (see Eq.~\ref{Eq:Grow}), 
\begin{eqnarray}
\mathcal{P}^{\textrm{gen}}_{ {\Gamma\!}_f} &=& { \chi_{V_m}({ {\Gamma\!}}_f)  \over \Omega_0^{ N_{{\Gamma}}+1}}{ 1\over p(r_{S,E},L_{{{S,E}}})} \frac{\exp[-\beta \; {\cal U}({{\Gamma\!}}_f)]}{W_{{\Gamma\!}_{f}}},
\label{Eq:Gammaf}
\end{eqnarray}

\noindent
where, as before, $\chi_{V_m}({\Gamma\!}_f)$ restricts ${\Gamma\!}_f$ to lay inside $V_m$. ${\cal U}({\Gamma\!}_f)$ and $W_{{\Gamma\!}_f}$ are the interaction energy and Rosenbluth factor of the free chain.

From Eq.~\ref{Eq:Partfunc} the equilibrium probabilities for a chain section $\Gamma$ to be in a free and in a cross-linked configuration are given by,
\begin{eqnarray}
\label{Eq:Prob_f}
\mathcal{P}_{{\Gamma\!}_f} &=& \chi_{V_m}({{\Gamma\!}}_f) \exp[-\beta \; {\cal U}({{\Gamma\!}}_f)]
\\
\label{Eq:Prob_b}
\mathcal{P}_{{\Gamma}_b} &=&  
{ \chi_{V_m}({{\Gamma}}_b) \over \Omega_0}\;
{ q_{A_2} \over q_A^2} \;
\exp[-\beta \; {\cal U}({{\Gamma}}_b)]\;
\delta(r_{\alpha,\beta}-\sigma)
\end{eqnarray}

Given the growing procedure defined above and the detailed balanced {for the flowchart below} as derived in Appendix~\ref{App:DetBal}, the following acceptance criteria are obtained for a binding and unbinding attempt, 
\begin{widetext}
\begin{eqnarray}
\label{Eq:Acc_fb}
\mathcal{P}^{\textrm{acc}}_{f \rightarrow b} &=& \;
\min \left[1,  
K_{\textrm{eq}}\;
\frac{n_f \; n_{f,V_m}}{(n_{\nu}+1)}\;
\frac{p(r_{S,\alpha},L_{{S,\beta}}+1)\;p(r_{\beta,E},L_{{\beta,E}})}{p(r_{S,E},L_{{S,E}})}\;
\frac{W_{{\Gamma}_b}}{W_{{\Gamma}_f}}
\right] \\
\label{Eq:Acc_bf}
\mathcal{P}^{\textrm{acc}}_{b \rightarrow f } &=& \;
\min \left[1,  
\frac{1}{K_{\textrm{eq}}}\;
\frac{n_{\nu}}{(n_f+2)(n_{f,V_m}+1)}
\frac{p(r_{SE},L_{{S,E}})}{p(r_{S,\alpha},L_{{S,\beta}}+1)\;p(r_{\beta,E},L_{{\beta,E}})}
\frac{W_{{\Gamma\!}_f}}{W_{{\Gamma}_b}}
\right]
\end{eqnarray}
\end{widetext}

\noindent
where $n_f$ is the number of free (unbound) reactive monomers in the polymer, and $n_{f,V_A}$ is the number of free reactive monomers inside the sphere $V_m$.  Note that in the previous equations the Rosenbluth weights of the old configurations (appearing at the denominators of the r.h.s.\ terms) are obtained by a regrowth process. As usually done in configurational--bias methods \cite{Mooij1992,Siepmann1992} the (re)growth  process of old and new configurations follow identical steps, except for the generation of the $k$-th trial in the regrowing process, which is taken equal to the actual monomer ${\bf u}_i$.   
In the presence of dangles, the growth of monomers belonging to chain with unconstrained end--point ($\Gamma_{b2}$ and $\Gamma_{b1}$ in Fig.~\ref{Fig:Figure4} (b) and (c)) is not biased by the guiding function $p$ (see Eq.\ \ref{Eq:Gen_k}), but the $k$ trials are generated uniformly. Accordingly, the acceptance rules are also different and can be obtained from Eq.\ \ref{Eq:Acc_fb} and \ref{Eq:Acc_bf} by setting to 1 all end-to-end probability densities $p$ that are function of the missing end terminal.  
\\

\noindent
{\bf Flow chart of the algorithm}\\

\noindent
A binding or unbinding attempt is chosen randomly with equal probability. \\

\noindent
{\it  Binding}
\begin{enumerate}
	\item A reactive site $\alpha$ is chosen randomly within the $n_f$ unbound ones. If all reactive complexes are cross-linked the move is rejected.
	\item A second reactive site $\beta$ is chosen randomly from all non-reacted $n_{f,V_m}$ sites that are inside the sphere $V_m$.
	\item The subchain $\Gamma$ inside the sphere $V_m$ containing $\beta$ is identified. 
	\item The feasibility of cross-linking $\alpha$ with $\beta$ is checked, if the cross-linking growth is not feasible the trial move is rejected.
	\item The direction of growth, i.e. $S \rightarrow \alpha \rightarrow E$ or $E \rightarrow \alpha \rightarrow S$ is selected randomly with equal probability.
	\item A new cross-linked configuration ${\Gamma}_b$, where $\alpha$ and $\beta$ are linked, is generated following the two-stage fix-end growth procedure described previously. The probability of generating ${\Gamma}_b$ is given by Eq.~\ref{Eq:Gammab}. 
	\item A new cross-linked configuration ${\Gamma}_b$ is accepted with probability defined by Eq.~\ref{Eq:Acc_fb}. 
\end{enumerate}

\noindent
{\it  Unbinding}  
\begin{enumerate}
	\item A cross-linking complex is chosen randomly from all available $n_{\nu}$ complexes. If there are no cross-linked complexes available, the move is rejected.
	\item The labels $\alpha$ and $\beta$ are randomly assigned to the two monomers forming the chosen cross--linked complex. Accordingly, $E$ and $S$ are identified.
	\item The direction of growth, either $S\rightarrow E$ or $E \rightarrow S$, is randomly selected.
	\item The subchain containing the reactive site $\beta$ is grown between the limiting segments, $S$ and $E$, using the procedure described in Sec.~\ref{Sec:Grow}.
	\item A new free configuration is accepted with probability defined by Eq.~\ref{Eq:Acc_bf}.
\end{enumerate}

The previous algorithms have been written for chain sections of the type reported in Fig.\ \ref{Fig:Figure3}. In presence of dangles, the growth procedure is altered for the fraction of chains containing the end/start monomer.

\subsection{Relaxing chain backbones}\label{Sec:Relax}

Beyond considering topological moves that change the linking state of the chain, standard Monte Carlo moves are used to relax the polymer backbone at fixed connectivity matrices $\nu$. Non-local configurational changes are
obtained by a single--pivot move in which one side of the
chain is rotated around a randomly selected monomer by
a random angle around a randomly chosen orientation~\cite{frenkel2001understanding,Baschnagel2004}. 
Each monomer has an associated loop variable $\ell=0,1,\cdots$ counting the number of loops to which the monomer belongs. A single--pivot move is allowed only in the case that the loop variable of the selected monomer has a value equal to $\ell=0$ or $\ell=$1, the latter case proceeds only if the monomer itself is reacted. We also use double--pivot moves, in which a fraction of the polymer is rotated by a random angle around the axis joining two randomly selected monomers. The move is rejected if the selected monomers belong to different loops. Rotations in the single-- and double--pivot move are achieved using quaternions to maximize the performance of the algorithm.

Additionally, backbone relaxation is promoted by the regrowth of internal sections of the polymer as described in Sec.~\ref{Sec:Grow}.\cite{Dijkstra1994,Pant1995,Escobedo1995b,Vendruscolo1997,Wick2000,Uhlherr2000,Chen2000b,Sepehri2017a} Chain sections including one of the chain ends are regrown using the standard configurational bias Monte Carlo method~\cite{Mooij1992,dePablo1992,Siepmann1992}. The position of reacted complexes (let say made of monomer $\alpha$ and monomer $\beta$) is further relaxed by randomly displacing one of its reacted monomers (e.g.\ $\alpha$) and by regrowing first a faction of the chain containing $\alpha$, and then a fraction of the chain containing $\beta$ by using a two stage process that binds $\beta$ to $\alpha$. In this case the growth processes cannot be implemented by means of reactive spheres as done in Fig.\ \ref{Fig:Figure3} because both $\alpha$ and $\beta$ change position. Instead we regrow a fixed number of segments, eventually limited by the presence of reacted monomers.

\section{Validation of the algorithm}\label{sec:alg-val}

\begin{figure}
	\centering
	\includegraphics[width=0.45\textwidth]{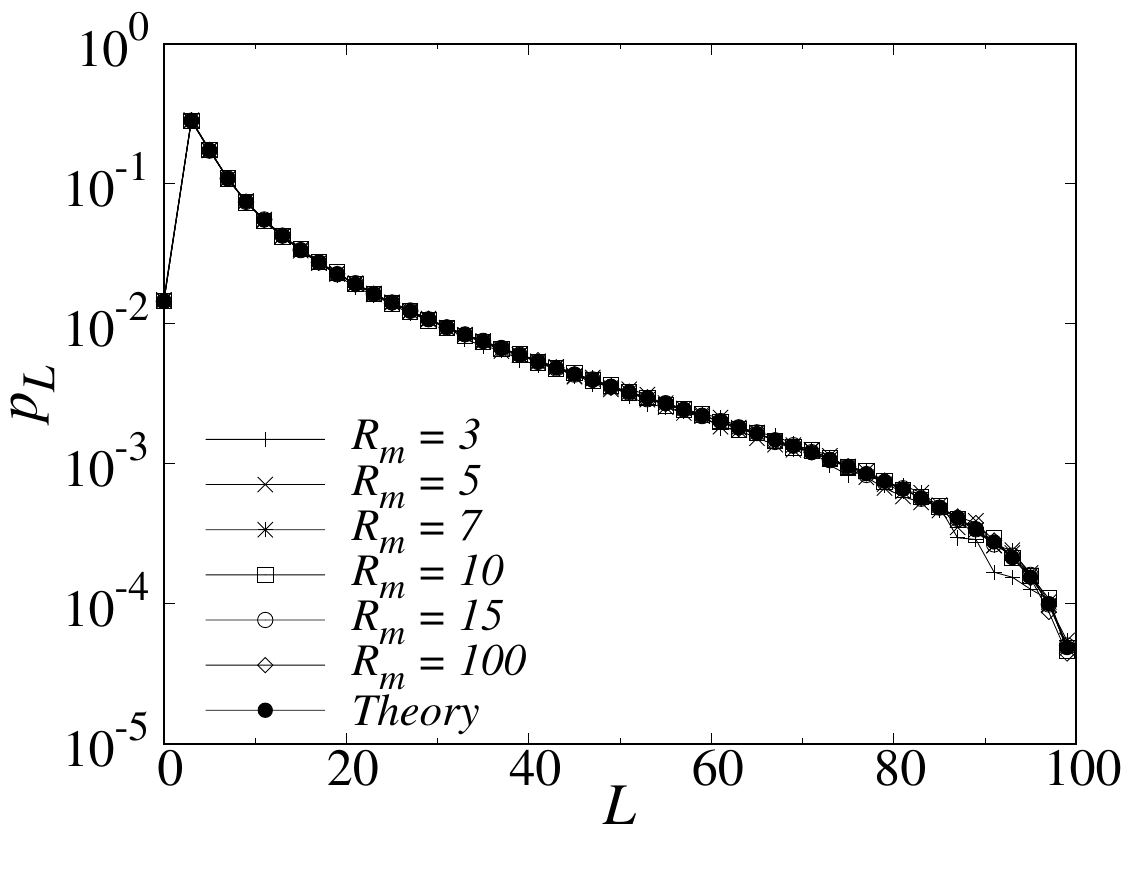}
	\caption{
		Relative probability $p_L$ of forming one loop of size $L$ for different reacting spheres specified by the radius $R_m$ (see Fig.\ \ref{Fig:Figure3}). $L{=}0$ corresponds to the no--loop configuration. Simulations results were obtained using a chain made of 100 monomers with 50\% of regularly distributed reactive monomers. We used $K_\mathrm{eq}{=}10$. Error bars are visible and larger than the symbols only for the smallest radius $R_m$ (in this work expressed in unit of $\sigma$). Theoretical values are calculated from probability distributions of freely--jointed chains made of $L$ segments and end-to-end distance equal to 0~\cite{Treloar1946,Yamakawa1971}.}
	\label{Fig:Figure5}
\end{figure}

\begin{table}
	\caption{Probabilities of forming configurations featuring up to two loops for freely-jointed chains made of 20 monomers with 4 equally spaced reactive sites and $K_\mathrm{eq}=100$. $p_{i,j}$ ($i,j>0$) refers to the probability of forming two loops starting from complexes separated by $i$ and $j$ segments. For compactness we have not listed 1-loop configurations but have reported relative probabilities conditional on having none ($p_{0,0}$) or two loops. Theoretical values are calculated using the expressions derived in Appendix \ref{App:Analytic}.}
	{\renewcommand{\arraystretch}{4}
		\begin{tabular}{c@{\quad} c@{\quad \quad} c@{\quad \quad} c}
			\hline
			\addlinespace[-2em]
			\multicolumn{2}{c}{{\bf Configuration}} & {\bf Theory} & {\bf Simulation}\\
			\addlinespace[-1em]
			\hline
			\addlinespace[-1em]
			$p_{0,0}$ \quad &
			\raisebox{-0.8\normalbaselineskip}{
				\includegraphics[scale=0.10]{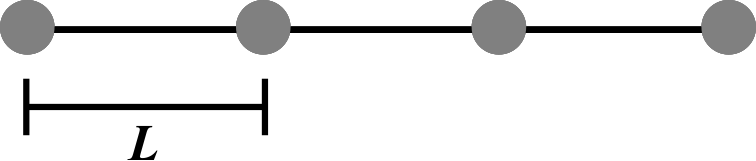}}& 
			0.1350 & 0.1347 $\pm$ 0.0004\\
			$p_{5,5}$ \quad &
			\raisebox{-0.8\normalbaselineskip}{
				\includegraphics[scale=0.10]{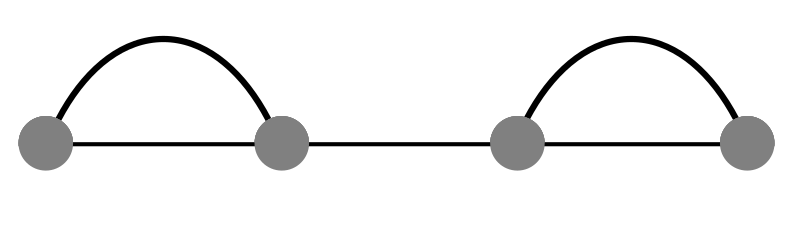}}& 
			0.5343& 0.5335 $\pm$ 0.0008\\
			$p_{20,5}$ \quad &
			\raisebox{-0.5\normalbaselineskip}{
				\includegraphics[scale=0.10]{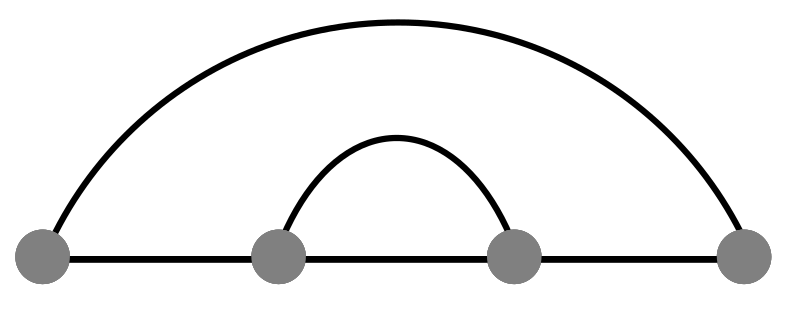}}& 
			0.2001& 0.2000 $\pm$ 0.0005\\
			$p_{10,10}$ \quad &
			\raisebox{-1.5\normalbaselineskip}{
				\includegraphics[scale=0.10]{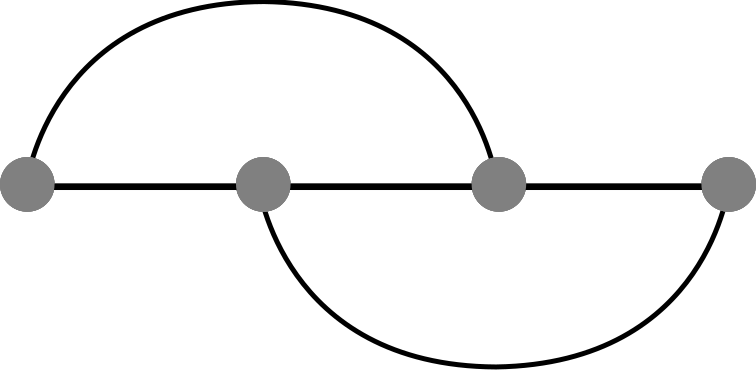}}& 
			0.1306& 0.1318 $\pm$ 0.0004\\[1em]\hline
		\end{tabular}
	}
	\label{Tab:Table1}
\end{table}	 
	
To validate our algorithm we consider freely--jointed chains (${\cal U}=0$ in Eq.\ \ref{Eq:Partfunc}), and run simulations in which we constrain the system to feature no more than one loop at the time. This is implemented by rejecting all attempts of forming new loops when one loop is already present in the system. Using Eq.\ \ref{Eq:Partfunc}, it can be shown that the ratio between the probability of not having any loops ($p_0$) and the probability of forming a single loop of length $L$ ($p_L$) is equal to $p_0/p_L=1/(K_\mathrm{eq} n_L)$, where $n_L$ is the number of loops of length $L$ featured by the chain. In Fig.~\ref{Fig:Figure5} we show that the theoretical predictions are perfectly matched by simulations. Importantly, simulations at different reacting radii, $R_m$ (see Fig.\ \ref{Fig:Figure5}), provide the same results. This is an important test that shows how entropic barriers of reacting complexes are properly accounted by our method.\\
We then run simulations allowing for configurations featuring two loops simultaneously. We consider an ideal chain functionalized by four $A$ reactive monomers. Three different single-- and two--loop configurations are possible. Tab.~\ref{Tab:Table1} shows only the relative probability of open chain and two--loop configurations, confirming that also in this case simulations and theoretical predictions are in agreement. As for the previous example different $R_m$ reproduce the same results.\\




\section{Morphology of reversibly cross-linked polymeric nanoparticles}
\label{Sec:Results}

\subsection{Model and simulation details}

We study the effect of reversible cross-linking on single-chain polymeric nanoparticles having different degrees of functionalization (number of reactive monomers) and affinity between reactive species (magnitude of the reaction constant $K_{\rm eq}$). Moreover, we study the effect of chain structure by considering random (RAN) and regular (REG) distributions of the reactive monomers in the chain. We consider ideal freely--jointed chains (ID) and chains in good solvent conditions in which monomers interact through a Weeks-Chandler-Andersen (WCA) potential,

\begin{align}
{\cal \beta \; U}(r_{ij}) = 
\left\lbrace
\begin{array}{l@{}c@{}l}
\displaystyle 4\left[\left(\frac{\sigma}{r_{ij}}\right)^{12} - \left(\frac{\sigma}{r_{ij}}\right)^{6}\right]+1
 &{}\quad;&  {} \quad r_{ij} \leq 2^{1/6}\sigma \\ [2em]
0 &{}\quad ;& {} \quad r_{ij} > 2^{1/6}\sigma
\end{array}
\right. 
\label{Eq:HS}
\end{align}
\noindent
where $r_{ij}$ is the distance between monomer $i$ and $j$. WCA potentials have already been used to study systems of functionalized polymers~\cite{LoVerso2015,Moreno2017}. In both cases (ID and WCA), chains are considered to be fully--flexible.

In all simulations we used precursors made of $N_C{=}100$ monomers. Different degrees of functionalization, defined as $f=N_R/N_C$, were studied with $f$ between 0 and 1. Initial configurations were generated by growing chains using a CBMC scheme~\cite{Mooij1992,dePablo1992,Siepmann1992}. We used production runs made by $5\times10^6$ MC cycles that were started after an equally long equilibration period. Each run took about 5 hours in a 2.3 GHz processor. In each MC cycle one of the following trial moves was selected with equal probability: single--pivot, double--pivot, chain regrowth, displacement of reacted complexes, and linking or unlinking of reactive monomers. Final results were typically averaged using 50 independent runs (and accordingly 50 different chains in the case of RAN functionalization). 
{ We report results in units of $\sigma$ and $k_B T$. In particular, the equilibrium constant, $K_\mathrm{eq}$, is expressed in unit of $\sigma^3$.}

\subsection{Structural properties of cross-linking polymers}

We study the effect of the degree of functionalization ($f$), organization of the reacting sites (REG or RAN), and strength of the linkages ($K_\mathrm{eq}$) on the morphology of polymeric particles. In particular, we consider the radius of gyration square $\left<R_g^2\right>$, where the brackets refer to the average taken over the ensemble defined by Eq.\ \ref{Eq:Partfunc} and over different chain realizations for RAN chains. $\left<R_g^2\right>$ is given by the principal moments of inertia of the chain \cite{flory1953}, 
\begin{eqnarray}
\left <R_g^2 \right > = \langle \Lambda_1 \rangle + \langle \Lambda_2 \rangle + \langle \Lambda_3 \rangle,
\end{eqnarray}
\noindent
where $\langle \Lambda_1 \rangle$, $\langle \Lambda_2 \rangle$, and $\langle \Lambda_3 \rangle$ are the eigenvalues of the gyration tensor. We also consider the average number of cross--linked complexes $\langle n_\nu \rangle$.

\begin{figure}
	\centering
	\includegraphics[width=0.5\textwidth]{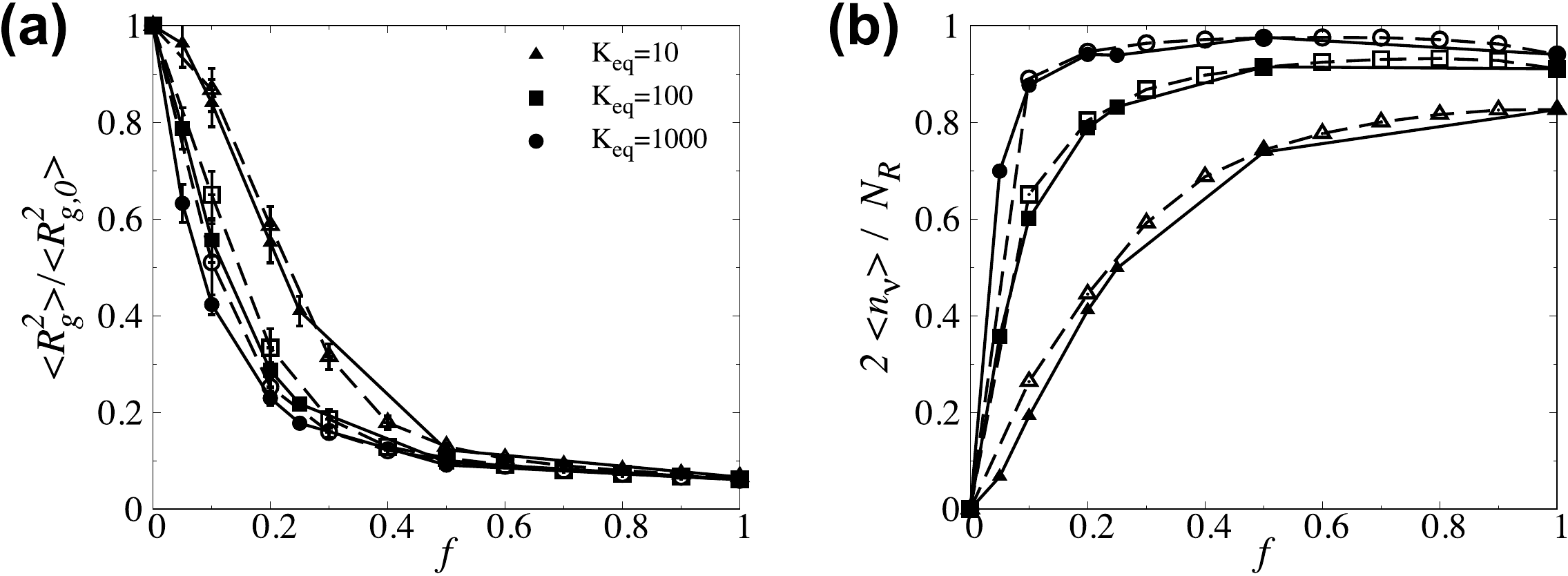}
	\caption{
		Radius of gyration $\left<R^2_g\right>$ and fraction of bound reactive monomers for ideal chains as a function of the functionalization degree $f$ for different magnitudes of the equilibrium constant $K_{\rm eq}$. REG (solid lines) and RAN (broken lines) functionalized chains are compared. The radius of gyration is expressed relative to the radius of gyration of the precursor ($f{=}0$) $\left<R^2_0\right>=\sigma^2 N_C/6 {=}16.67\cdot \sigma^2$.}
	\label{Fig:Figure6}
\end{figure}

Fig.~\ref{Fig:Figure6} shows results for $\langle R_g^2 \rangle$ and $\langle n_\nu \rangle$ for freely--jointed chains as a function of the degree of functionalization at different $K_\mathrm{eq}$. Larger values of $f$ reduces the size of the molecule until reaching a plateau for $f{\approx}0.4$. Similar trends have been reported previously~\cite{Moreno2017,LoVerso2014,Pomposo2014,Stals2014,Pomposo2017}, although a direct comparison is not possible in view of the reversible nature of our linkages. 
The fraction of reacting sites forming cross-linked complexes approaches unity as $f$ and $K_\mathrm{eq}$ increase. The limiting values for the radius of gyration and the fraction of reacting sites are explained by the difficulty of forming states with longer loops due to higher configurational costs. Regularly and randomly functionalized chains behave similarly, with a slight reduction of the molecular size for REG systems compared to the RAN case. Such discrepancy is more evident at low degrees of functionalization.

\begin{figure}
	\centering
	\includegraphics[width=0.5\textwidth]{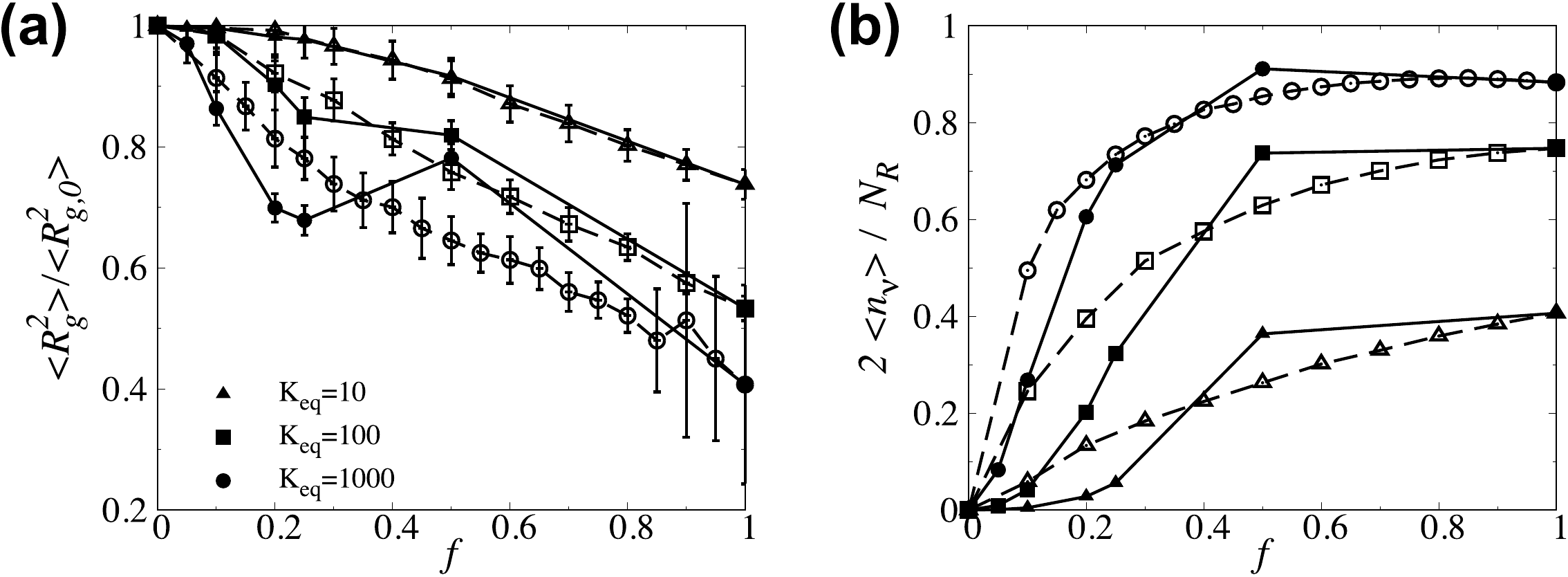}
	\caption{
		 Radius of gyration $\left<R^2_g\right>$ and fraction of bound reactive sites for WCA chains as a function of functionalization degree $f$ for REG (solid lines) and RAN (broken lines) chains. The radius of gyration of the precursor ($f{=}0$) is equal to $\left<R^2_0\right>{=}57.20\cdot \sigma^2$ as obtained from simulations. The notation is identical to what used in Fig.~\ref{Fig:Figure6}.
		 }\label{Fig:Figure7}
\end{figure}

Gyration radii and number of cross--linked complexes of equilibrated WCA chains are shown in Fig.~\ref{Fig:Figure7}. In contrast to the ideal case, at high $K_\mathrm{eq}$, $\langle R^2_g\rangle$ is re--entrant in $f$. In particular, the size of the $f{=}0.5$ REG chain is bigger than the $f{=}1$ and $f{=}0.25$ REG chains. The plots of $\langle R^2_g \rangle$ are mirrored by the average number of reacted complexes $\langle n_{\nu} \rangle$ that, for REG chains, is first below and then above the results obtained using RAN chains. The re--entrant behavior can be explained by an increasing competition between chain compaction, promoted by larger values of $f$ and $K_{\rm eq}$, and chain swelling due to excluded volume interactions. The re-entrant behavior is not observed in chains with randomly distributed reactive monomers due to different loop length distributions, as highlighted below. Morphological differences between ideal and WCA chains are mainly due to the fact that excluded volume interactions favors short loop conformations, as it has been extensively highlighted previously by other authors \cite{Moreno2016,Moreno2013,LoVerso2014,terHuurne2017,Stals2014,Pomposo2014}. Here we corroborate this general scenario but warn that peculiar features, like the re--entrant behavior of Fig.\ \ref{Fig:Figure7}, may be difficult to detect in the irreversible-limit as shown in Sec.\ \ref{Sec:Irreversible}.

\begin{figure}
	\centering
	\includegraphics[width=0.4\textwidth]{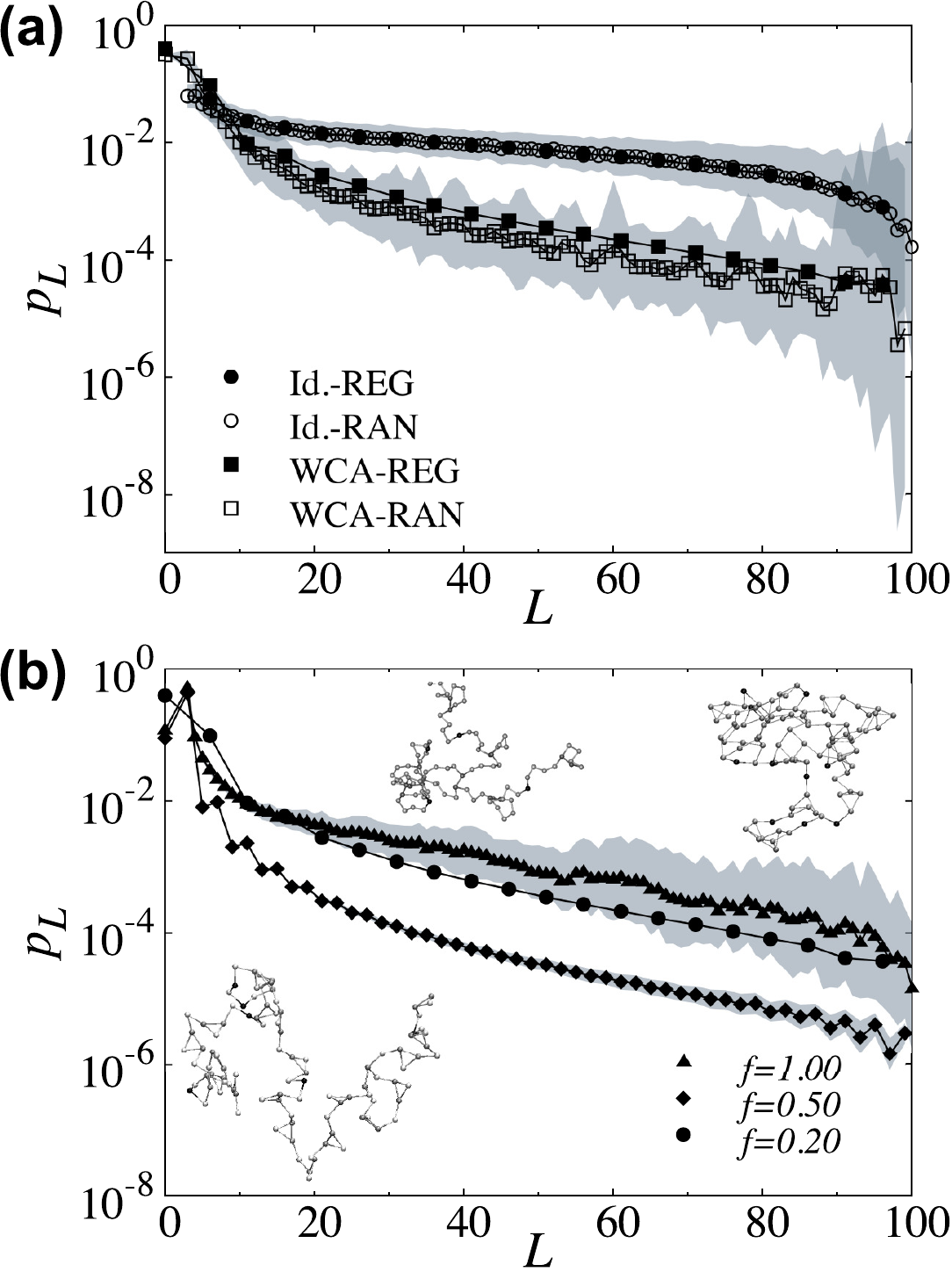}
	\caption{ Loop length distributions as a function of the type of functionalization (a), and the degree of functionalization (b). Data points at $L=0$ are the fraction of unbound complexes (see Figs.~\ref{Fig:Figure6} and \ref{Fig:Figure7}). In all cases $K_\mathrm{eq}=1000$ while in (a) $f=0.2$. In (b) we considered REG functionalizations. Snapshots correspond to $f=0.2$ and 0.5 (bottom) and $f=1$ (top). Error bars are smaller than the symbols in the case of REG chains. For RAN chains we reported a confidence interval equal to one standard deviation (shaded area). In RAN systems the variance is dominated by the quenched distribution of reactive monomers.
	}
	\label{Fig:Figure8}
\end{figure}

\begin{figure}
	\centering
	\includegraphics[width=0.5\textwidth]{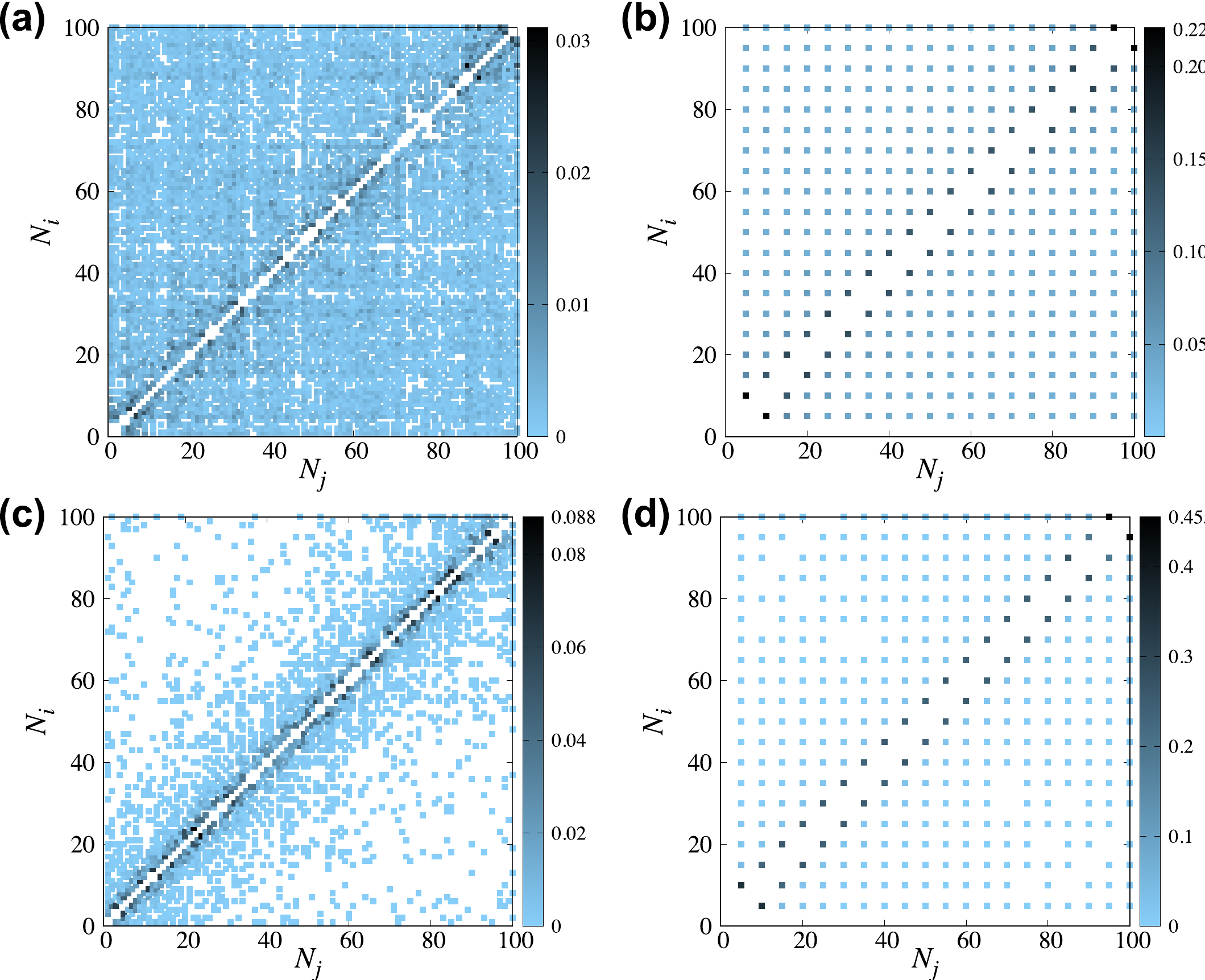}
	\caption{
		Average connectivity maps $\left<\nu_{ij}\right>$ (see Fig.\ \ref{Fig:Figure1}) for chains with a degree of functionalization $f{=}0.2$ and $K_{\rm eq}{=}1000$. The colors on the heat map indicate the probability of forming a loop. The upper maps, (a) and (b), correspond to ideal chains and the bottom maps, (c) and (d), to WCA chains. Maps on the left column, (a) and (c), were obtained from randomly functionalized chains. Maps on the right column, (b) and (d), were obtained from regularly functionalized chains.
	}
	\label{Fig:Figure9}
\end{figure}

Loop statistics is studied in Figs.~\ref{Fig:Figure8} and \ref{Fig:Figure9}. Fig.~\ref{Fig:Figure8}(a) shows the probability distributions of having loops of length $L$ for freely--jointed and WCA chains at $f{=}0.2$ and $K_{\rm eq}{=}1000$ (corresponding to a higher compaction degree of RAN as compared to REG chains for the non--ideal system, see Fig.\ \ref{Fig:Figure7}). 
Interestingly, the loop length distributions of RAN and REG ideal chains nicely overlap. This explains why the gyration radius in the two cases is similar (see Fig.~\ref{Fig:Figure6}). Such agreement disappears in the case of WCA chains where RAN functionalizations enhance the presence of short loops at the cost of a lower relative probability of forming longer loops compared to REG chains. This results in chains that are more compact in the REG case (see Fig.~\ref{Fig:Figure7}).
Note that short loops are also present in REG ideal chains. However because short loops are less dominant in ideal conditions such effect is not sufficient to differentiate the size of the cross--linked chain. { 
Fig.~\ref{Fig:Figure8}(b) studies REG functionalizations with $f=0.2$, 0.5, and 1, and clarifies the two driving mechanisms leading to chain compaction: loop length and amount of reacted complexes. On the one side, the $f=0.2$ chain is smaller than the $f=0.5$ one because of longer loops and larger scale branching. On the other, the $f=1$ chain is smaller than the $f=0.5$ one because of more reacting monomers leading somehow to the formation of very long loops. The interplay between these different morphologies deserves further investigation.
}

To further clarify the statistics of cross--linked complexes in Fig.~\ref{Fig:Figure9} we report averaged connectivity maps for $f{=}0.2$ and  $K_{\rm eq}{=}1000$. { Panels (a), (c)} and { (b), (d)} refer, respectively, to ideal and WCA chains, while { panel (a), (b) and (c), (d) } to RAN and REG chains.
The color of the heat map reflects the probability that a determined loop is visited during the simulation. From the connectivity maps of ideal chains, it can be observed that our method is able to sample all possible pairs of reacted complexes, validating and justifying the scheme. The connectivity map for WCA chains confirms the results of Fig.\ \ref{Fig:Figure8} and shows that long loops are rarely formed. The regular distribution of reactive monomers avoids, as stated before, the formation of short loops resulting in smaller chain nanoparticles.
{ For REG-WCA chain, see Fig. 9d, monomer 100 binds more often than monomer 5. Such asymmetry is due to steric repulsions engendered by the four inert beads flanking monomer 5. A study about hybridization of DNA strands featuring inert tails \cite{DiMichele2014} reported similar results. 
}

\begin{figure}
	\centering
	\includegraphics[width=0.5\textwidth]{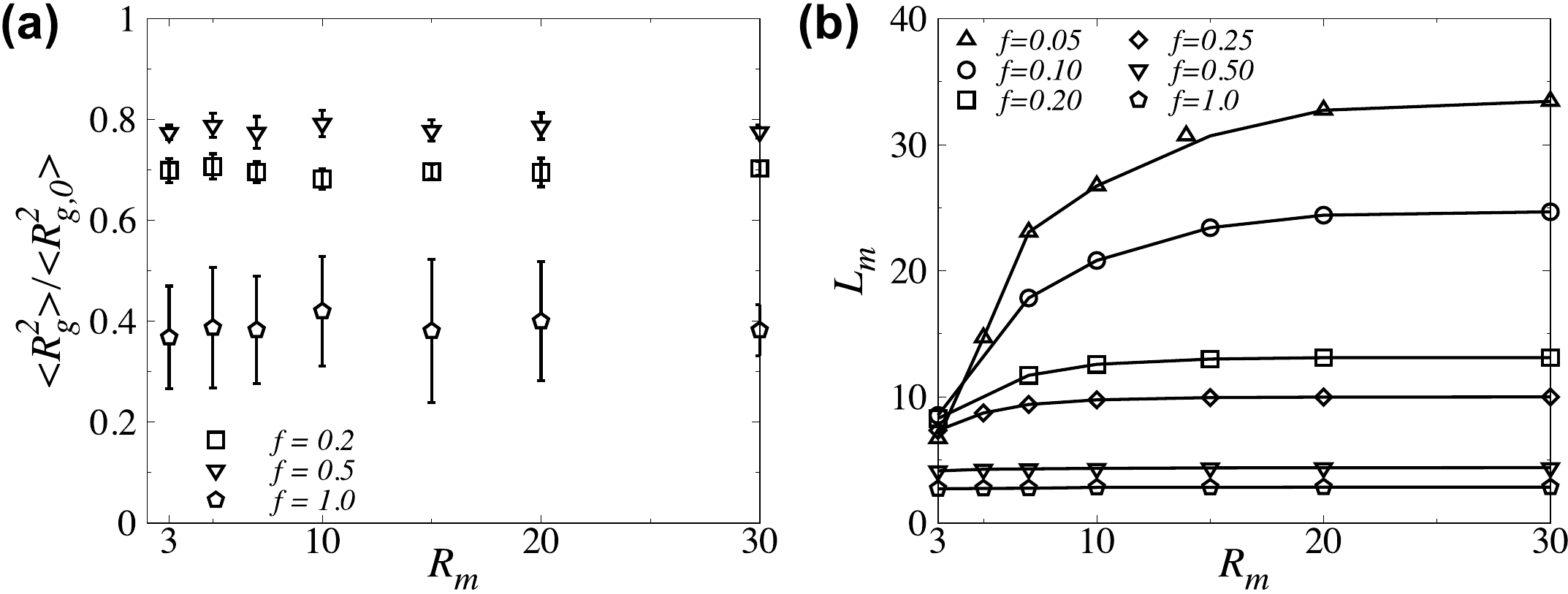}
	\caption{
	 (a) Radius of gyration $\left<R^2_g\right>$ as a function of the radius of the reactive sphere $R_m$ for REG-WCA chains at $K_{\rm eq}{=}1000$ and different degrees of functionalization $f$. 
	 {
	 (b) Average length of the grown chain section $\Gamma$ (see Fig.~\ref{Fig:Figure3}) for REG-WCA chains at $K_{\rm eq}{=}1000$ as a function of $R_m$.
	 }}
	\label{Fig:Figure10}
\end{figure}

\begin{figure}
	\centering
	\includegraphics[width=0.5\textwidth]{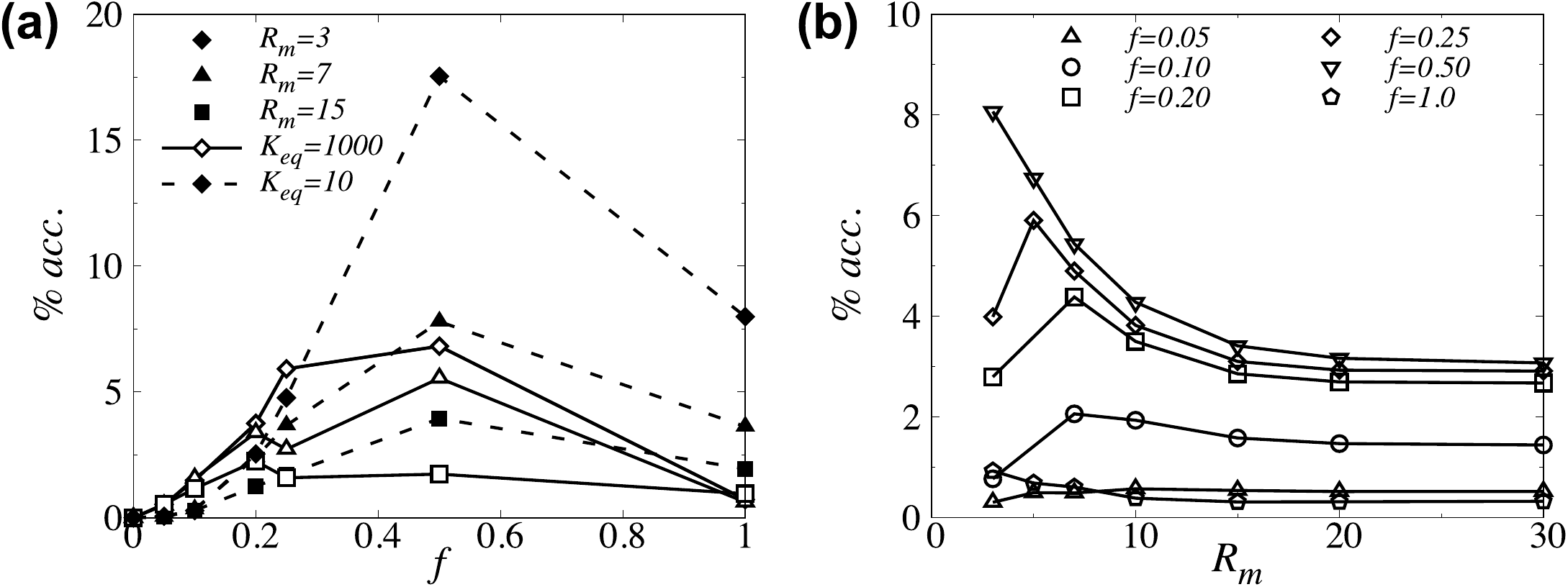}
	\caption{
	Acceptance of the binding/unbinding move (a) as a function of the degree of functionalization, $f$, (b) and of the reactive radius, $R_m$. In (b) $K_\mathrm{eq}=1000$.
	 }
	\label{Fig:Figure10b}
\end{figure}

{ \subsection{Reacting radius and efficiency}

In this section we study how $R_m$ affects simulations. 
First, we validate our algorithm by performing simulations at various values of $R_m$. Fig.~
\ref{Fig:Figure10} (a) confirms that results obtained using different reactive radii are equivalent within statistical error. In Fig.~\ref{Fig:Figure10b} we compare the acceptance of the linking/unlinking move as a function of $R_m$, $K_\mathrm{eq}$, and $f$. 
Fig.~\ref{Fig:Figure10b} (a) shows that the algorithm performs better at intermediate values of $f$. At high values of $f$, free reactive monomers are more likely to be close to some dimers (limiting the length of the segment $\Gamma$), and linking becomes difficult given the finite extensibility of the chain. This result is confirmed by the fact that, at high $f$, acceptances are higher for lower values of $K_\mathrm{eq}$ when fewer loops are present. Fig.~\ref{Fig:Figure10b} (a) shows that acceptances decrease at low values of $f$. In this limit, free reactive monomers become scarce and the growth of long segments is more difficult.  Fig.~\ref{Fig:Figure10} (b) studies the correlation between $R_m$ and the average length of the grown segment, $L_m$. At low values of $f$, $L_m$ increases until reaching a plateau for values of $R_m$ comparable with the size of the chain. As observed before, at high values of $f$, $L_m$ is not affected by $R_m$ given that 
the growth of long loops is rarely attempted. \\ 
A possible source of inefficiency is related to the fact that we are using as guiding functions end-to-end distances of ideal, rather than WCA, chains. The use of ideal end-to-end distributions is not optimal when generating long-chain segments. We quantify such effect in Fig.~\ref{Fig:Figure10b} (b), where we plot acceptances as a function of $R_m$. For intermediate values of $f$, the maximum values of the acceptance are between $R_m=5$ and $R_m=10$. Acceptances slightly decrease at larger values of $R_m$. This tiny reduction may be symptomatic of inefficiencies related to bad guiding functions. However, Fig.~\ref{Fig:Figure10} (b) also shows that for $R_m=5-10$ the maximum values of $L_m$ has already been reached in all systems. It is quite likely than that inefficiencies at large $L_m$ are mainly due to the finite extensibility of the segments. 
}

\subsection{Irreversible cross-linking}\label{Sec:Irreversible}

\begin{figure}
	\centering
	\includegraphics[width=0.5\textwidth]{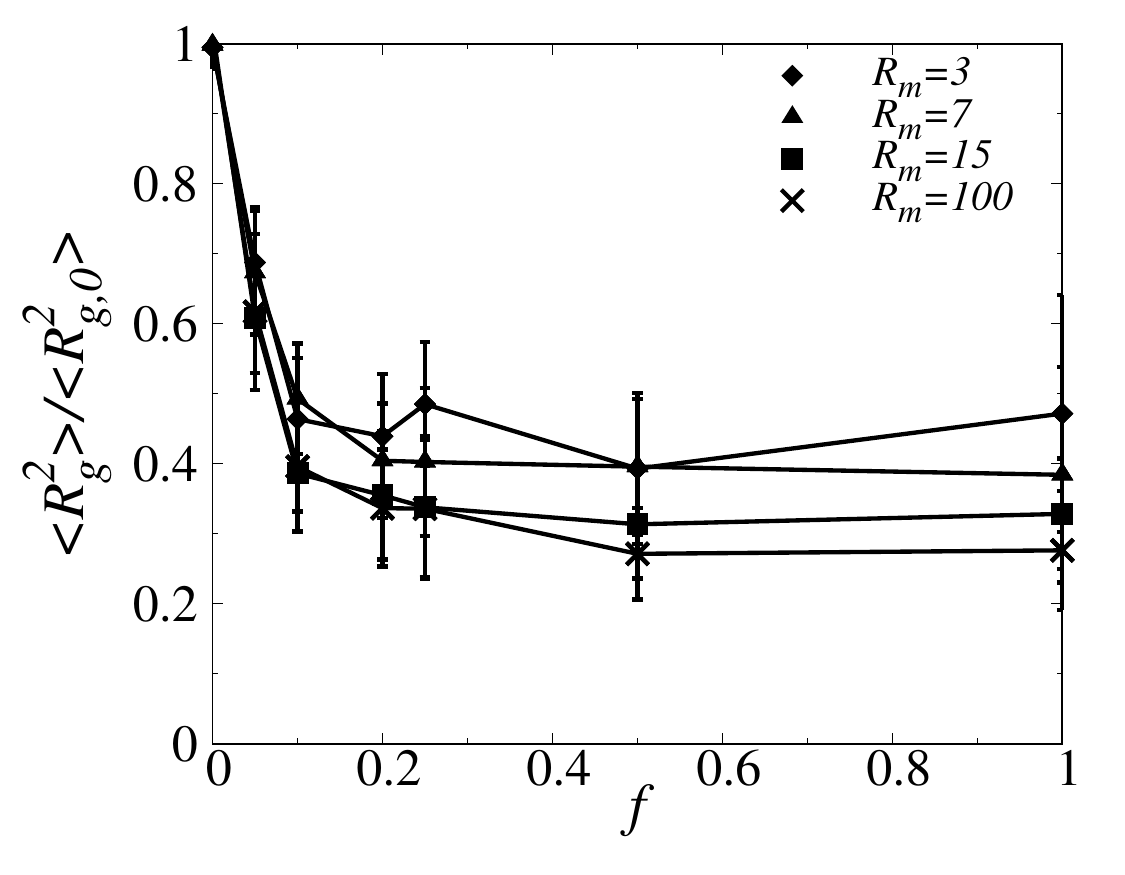}
	\caption{
	 Radius of gyration $\left<R^2_g\right>$ as a function of functionalization degree $f$ for REG-WCA chains in the irreversible limit, for different values of $R_m$. }
	\label{Fig:Figure11}
\end{figure}

We also studied cross-linking in the irreversible limit ($K_{\rm eq}{\rightarrow} \infty$). Such limit was reproduced by rejecting any unlinking attempt (see Fig.~\ref{Fig:Figure3}). Results for the gyration radius are reported in Fig.~\ref{Fig:Figure11}. Interestingly, Fig.~\ref{Fig:Figure11} shows that results obtained using different reacting radius $R_m$ do not agree. 
This highlights the fact that the system fails to equilibrate because, as we discussed in the previous sections, equilibrium sampling is not affected by $R_m$ in view of the reversibility of the algorithm.  
We anticipate that equilibration can be reached also in the irreversible limit by employing a different topological move capable of swinging reactive sites between cross-linked complexes in a single step. This study will be presented elsewhere. 
For the purpose of this paper we observe how Fig.\ \ref{Fig:Figure11} proves that qualitative differences are expected in the morphology of SCPNs folded using irreversible (Fig.\ \ref{Fig:Figure11}) or reversible (Fig.\ \ref{Fig:Figure6} and \ref{Fig:Figure7}) linkages.

\section{Conclusions}
\label{Sec:Conc}

Polymeric networks forming supramolecular contacts are interesting systems because they offer the possibility of controlling intra-- and inter--molecular interactions starting from a given functionalization of the precursor specified by the strength, amount, and organization of the chemical groups along polymer backbones. 
Systems in which the amount of reacted complexes can be tuned by control parameters such as temperature or solvent conditions, allow for a remote control over molecular interactions, therefore providing the possibility to embed new responsive behaviors into the material. 
Designing reversible polymer networks featuring specific functionalities is difficult because large--scale properties of these materials are the result of tight competitions between chemical association of reactive sites and entropic contributions of the polymeric backbones. Such terms cannot be easily estimated unless using quantitative methodologies. 

Simulations of reversible supramolecular networks are scarce. Most of the available literature has focused on the irreversible limit in which, once a reaction happens, complexes are permanently linked \cite{Moreno2017,Moreno2013,LoVerso2014,LoVerso2015,Bae2017}. This hampers the study of functional properties (for instance self--assembly, resilience, or self--healing) that are based on many binding--unbinding events. The difficulty of simulating reversible flexible networks can be ascribed to large entropic barriers that need to be overcome to let the complexes react. Properly accounting of reactions has been typically achieved by a detailed (often quantum mechanical) description of the species involved in the reaction. This approach becomes unpractical when attempting to sample polymer backbones functionalized with reactive sites. 

In this study we proposed a methodology that allows sampling supramolecular networks using coarse--grained models and non--local Monte Carlo moves. We use a description in which only the features of the reacting complex controlling their interaction with the backbones, like size or shape of the reacted dimer, are explicitly retained. Chemical details are modeled by means of internal partition functions that are linked to the equilibrium constant of free complexes in solution. The latter can be calculated using quantum-chemistry methods or estimated by experiments. 
Reactions between complexes are implemented using configurational bias moves powered by the possibility of changing the topology of the network. Reaction within complexes are constrained to happen within a reaction volume $V_m$ (of radius $R_m$). For instance, when attempting to react two complexes, the backbone of the network inside $V_m$ is regrown in a way to constrain two reacting sites to be at the exact relative position that allows for the formation of a reversible linkage, while preserving the connectivity of chain backbones at the boundary of $V_m$. The reverse move is designed to satisfy detailed balance also accounting for chemical contributions entering the expression of the acceptance rules {\em via} the chemical equilibrium constant of free complexes in solution. The size of the interacting sphere determines the efficiency of the algorithm: For large $R_m$ the acceptance decreases because binding far away complexes becomes more difficult. On the other hand for small $R_m$ the dynamics of the algorithm resembles physical dynamics. In the latter case reactions are limited by the time taken by pairs of reactants to diffuse at relative distances smaller than $R_m$.
{ 
The linking/unlinking algorithm does not prevent bond crossing. This problem is relevant in the presence of irreversible linkages and underlies all non—local Monte Carlo moves including pivots and CBMC regrowths. Intriguingly, the reactive sphere $V_m$ may be used to design checks to enforce topological constraints. Let consider, for instance, the ring $R_1$ obtained by joining the old and the new configuration of a linking attempt ($\Gamma_f$ and $\Gamma_b$ in Fig.~\ref{Fig:Figure3}). By calculating topological integrals \cite{Micheletti2011} or by using shrinking algorithms \cite{Padding2001}, it is possible, within certain approximations, to detect trial moves that may change the topology of the network.
}

In this work we have applied the proposed algorithm to the study of self--assembly of single chain polymeric nanoparticles (SCPNs). In these systems a single chain backbone is functionalized by reacting complexes that drives the folding of the chain by forming intra--molecular linkages. It has been extensively highlighted how SCPNs often resemble proteins with intrinsically disordered regions showing large conformational fluctuations \cite{Moreno2016}. Simulation studies on SCPNs have mainly focused on finding suitable protocols to relax the precursor resulting in compact structure when activating irreversible linkages \cite{Perez-Baena2014,Formanek2017,LoVerso2015}. Instead in this study we have highlighted differences in the morphology of SCPNs when assembled by means of reversible or irreversible linkages. First we have verified that reversible linkages allow sampling between many different sets of linkages including the possibility of forming short-lived long loops. This highlights the dynamic structure of these materials. We then concentrate on the difference between random and regular functionalization. As expected the average number of reacted complexes increases with the degree of functionalization in both cases. Surprisingly, for regular functionalization the size of the SCPNs is re-entrant: more compact particles are found at intermediate degrees of functionalization. This is because in good--solvent conditions short loops are favored to minimize excluded volume interactions. This effect disappears for randomly functionalized chains, where short loops are recurrent even at intermediate degrees of functionalizations, and for ideal chains. In the latter case the chains are more entangled resulting in average connectivity matrices that are more spread than for interacting chains. Intriguingly this re--entrant behavior disappears in the irreversible limit, in our simulations emulated by sampling using the linking algorithm as sole topological move.

As future perspective it will be interesting to study the effects of considering complexes constraining in different ways the reacted backbones and their effect on the large scale morphology of the SCPNs. Moreover it will be also important to extend the present study to multi--chain systems.

\section*{Acknowledgements}

Financial support was provided by the {\em F\'ed\'eration Wallonie-Bruxelles (Actions
de Recherches Concert\'ees)} for the project `Numerical design of supramolecular
interactions'. Computational resources have been provided by the Consortium
des Equipements de Calcul Intensif (CECI), funded by the Fonds de la
Recherche Scientifique de Belgique (F.R.S.-FNRS) under Grant No. 2.5020.11.

\appendix

\section{Internal partition functions and equilibrium constant of complexes free in solution}
\label{App:reac}

In this section we calculate the equilibrium constant $K_\mathrm{eq}$ of the reactive groups/species in solution in diluted conditions. In this limit intermolecular interactions between reacting species can be neglected. The partition function of $C$ species in solution is then given by~\cite{Johnson1994,Chen2000}, 

\begin{eqnarray}
Z^{\textrm{id}}(N_i,V,T) = \prod_{i=1}^{C}{z_i(V,T)^{N_i} \over N_i!} \, ,
\end{eqnarray}
\noindent
where $z_i$ and $N_i$ are the single molecule partition function and number of molecules of species $i$, respectively. The chemical potential of species $i$ can then be written as,
\begin{eqnarray}
\nonumber
{\mu_i(V,T) \over k_BT} = - {\partial \ln Z^{\textrm{id}} \over \partial N_i} \Big{|}_{N_{j \ne i},V,T} 
= - \ln {z_i(V,T) \over N_i} \, .
\label{Eq:Muideal}
\end{eqnarray}
Using the previous equation along with the chemical equilibrium condition $\sum_{i=1}^C \nu_i \mu_i = 0$, where $\nu_i$ are the stoichiometric coefficients of the reaction, we obtain the following relation between the equilibrium constant of the reaction and single molecule partition functions, 
\begin{eqnarray}
K_{\textrm{eq}} = \prod_{i=1}^C \rho_{i}^{\nu_i} =  \prod_{i=1}^C \left( {z_{i} \over V} \right)^{\nu_i},
\label{Eq:Keq}
\end{eqnarray}
where $\rho_i$ is the molar density of species $i$. We now consider the dimerization reaction $2A\leftrightharpoons A_2$, using the model introduced in Sec.\ \ref{Sec:Model} in which $A$ sites and $A_2$ complexes are modeled as single beads and dumbbells respectively. The single molecule partition function for each species can then be written as,
\begin{eqnarray}
z_A = V  \; q_A  & \qquad ; \qquad &
z_{A_2} =  {V \over 2}  \; q_{A_2},
\end{eqnarray}
\noindent
where $q_A$ and $q_{A_2}$ are internal partition functions that also include all momentum contributions, in particular, De Broglie thermal wavelengths if complexes are treated classically. The term 1/2 is included due to the indistinguishability of both monomers forming the dimer. Finally adapting Eq.\ \ref{Eq:Keq} to the dimerization reaction considered here we can relate the internal partition functions with the equilibrium constant as following,
\begin{eqnarray}
K_{\textrm{eq}} = \frac{q_{A_2}}{2 \; q_{A}^2} \, .
\label{Eq:Keq_q}
\end{eqnarray}
The previous relation has been used to parametrize the acceptance rules (see Eqs.~\ref{Eq:Acc_fb} and~\ref{Eq:Acc_bf}) using the equilibrium constant. $K_{\textrm{eq}}$ can be obtained by means of experiments (using Eq.~\ref{Eq:Keq}) or quantum mechanical/atomistic calculations (using  Eq.~\ref{Eq:Keq_q}). 
{
For instance, if we consider the case of two complexes interacting via a classical potential $V_{AA}$, we have
\begin{eqnarray}
K_{\textrm{eq}} = {\int \mathrm{d} \{ r_1\}  \mathrm{d} \{ r_2 \} e^{ -\beta [V_A(\{ r_1 \}) +  V_A(\{ r_2 \}) + V_{AA}(\{ r_1 \}, \{r_2\})] } \over
2 \big[ \int \mathrm{d} \{ r_1\} e^{-\beta V_A(\{r_1\})}\big]^2 /V
}
\nonumber \\
\end{eqnarray} 
where $\{ r_1 \}$ and $\{ r_2 \}$ are the atomistic variables of two $A$ molecules and $V_A$ is the set of intra-molecular interactions. 
}

\section{Detailed Balance for the binding/unbinding move}
\label{App:DetBal}

Here we derive the acceptance rules (Eqs.~\ref{Eq:Acc_fb} and \ref{Eq:Acc_bf}) for the binding/unbinding move defined by the algorithms detailed in Sec.\ \ref{Sec:Link}. Detailed balance condition between a bound ($b$) and a free ($f$) state is written as (see Fig.~\ref{Fig:Figure3} and Sec.\ \ref{Sec:Link} for the definitions and the notation used), 

\begin{widetext}
\begin{eqnarray}
\mathcal{P}_{b} \;
\mathcal{P}_{\Gamma_f} \;
\mathcal{P}_{{\alpha,f}} \;
\mathcal{P}_{{\beta},V_m} \;
\mathcal{P}_{\Gamma_b}^{\textrm{gen}}  \;
\mathcal{P}^{\textrm{acc}}_{f \rightarrow  b} \; = 
\mathcal{P}_{f} \;
\mathcal{P}_{\Gamma_b} \;
\mathcal{P}_{\alpha,\beta} \;
\mathcal{P}_{\alpha}
\mathcal{P}_{\Gamma_f}^{\textrm{gen}} \;
\mathcal{P}^{\textrm{acc}}_{b \rightarrow f}
\label{Eq:DetBal_fb}
\end{eqnarray}
\end{widetext}
\noindent
where, for the binding move,
\begin{longtable}{l l p{.4\textwidth}}
	$\mathcal{P}_{b}$ &:& Probability of performing a binding attempt, 1/2.\\
	$\mathcal{P}_{{\Gamma\!}_f}$ &:& Probability for the chain section $\Gamma$ to be in the actual free configuration ${\Gamma\!}_f$, Eq.~\ref{Eq:Prob_f}.\\
	$\mathcal{P}_{\alpha,f}$ &:& Probability of selecting a reactive monomer $\alpha$ from all unbound reactive monomers in the polymer, $1/(N_R - 2 n_{\nu}) = 1/n_f$.\\
	$\mathcal{P}_{\beta,V_m}$ &:& Probability of choosing an unbound reactive monomer $\beta$ inside the sphere $V_m$, $1/n_{\textrm{u},V_m}$.\\
	$\mathcal{P}^{\textrm{gen}}_{\Gamma_b}$ &:& Probability of generating a cross-linked configuration for $\Gamma$ by joining monomer $\alpha$ with $\beta$, Eq.~\ref{Eq:Gammab}.\\
	$\mathcal{P}^{\textrm{acc}}_{f \rightarrow b}$ &:& Acceptance probability for a configurational change of $\Gamma$ from a free to a bound state.\\
	$\mathcal{P}_{f}$ &:& Probability of performing an unbinding attempt, 1/2.\\	
	$\mathcal{P}_{\Gamma_b}$ &:& Probability for the chain section $\Gamma$ to be in a bound configuration $\Gamma_b$, Eq.~\ref{Eq:Prob_b}.\\
	$\mathcal{P}_{\alpha,\beta}$ &:& Probability of choosing a cross-linking complex containing $\alpha$ and $\beta$ to be unbound, $1/(n_{\nu}+1)$.\\
	$\mathcal{P}_{\alpha}$ &:& Probability of choosing $\alpha$ in the previously selected cross-linked complex, 1/2.\\
	$\mathcal{P}^{\textrm{gen}}_{\Gamma_f}$ &:& Probability of generating an unlinked configuration for $\Gamma$ by by unbinding $\alpha$ from $\beta$, Eq.~\ref{Eq:Gammaf}.\\	
	$\mathcal{P}^{\textrm{acc}}_{b \rightarrow f}$ &:& Acceptance probability for a configurational change of $\Gamma$ from a bound to a free state.\\
\end{longtable}
Considering the Metropolis scheme \cite{Metropolis1953}, a trial change from a free to a bound configuration is then accepted with probability,
\begin{eqnarray}
\nonumber
\mathcal{P}^{\textrm{acc}}_{f \rightarrow b} &=& \;
\min \left[1,  
\frac{\mathcal{P}_{\Gamma_b}}{\mathcal{P}_{{\Gamma\!}_f}}
\frac{\mathcal{P}_{\alpha,\beta}\mathcal{P}_{\alpha}}{\mathcal{P}_{\alpha,f} \mathcal{P}_{\beta,V_m}}
\frac{\mathcal{P}_{b \rightarrow f}^{\textrm{gen}}}
{\mathcal{P}_{f \rightarrow b}^{\textrm{gen}}} \right]
\end{eqnarray}
\noindent
Similarly the acceptance of an unbinding trial move is given by,
\begin{eqnarray}
\nonumber
\mathcal{P}^{\textrm{acc}}_{b \rightarrow f} &=& \;
\min \left[1,  
\frac{\mathcal{P}_{\Gamma_f}}{\mathcal{P}_{{\Gamma\!}_b}}
\frac{\mathcal{P}_{\alpha,f} \mathcal{P}_{\beta,V_m}}{\mathcal{P}_{\alpha,\beta}\mathcal{P}_{\alpha}}
\frac{\mathcal{P}_{f \rightarrow b}^{\textrm{gen}}}
{\mathcal{P}_{b \rightarrow f}^{\textrm{gen}}} \right]
\end{eqnarray}
\noindent
where in this case,
\begin{longtable}{l l p{.4\textwidth}}
	$\mathcal{P}_{\alpha,\beta}$ &:& Probability of choosing a cross-linking complex from all available ones, $1/n_{\nu}$.\\
	$\mathcal{P}_{\alpha,f}$ &:& Probability of selecting a reactive monomer $\alpha$ from all new free monomers in the polymer, $1/(n_f+2)$.\\
	$\mathcal{P}_{\beta,V_m}$ &:& Probability of choosing an unbound reactive monomer $\beta$ from all new free reactive monomers inside the sphere $V_m$, $1/(n_{f,V_m}+1)$.\\
\end{longtable}
\noindent

\section{Two--loop calculation of the partition function of ideal chains}\label{App:Analytic}
Here we report on the theoretical calculations relative to the second example presented in Sec.\ III E. We consider an ideal chain regularly functionalized by four complexes separated by $L$ segments. Using  Eqs.\ \ref{Eq:Partfunc} and \ref{Eq:Keq_q} the partition function of the system is written as, 
\begin{eqnarray}
{Z\over q_A^4}&=&1+ { K_\mathrm{eq} \over \Omega_0 } \left[
3 {\cal Z}_1+2{\cal Z}_2 + {\cal Z}_3
\right]
\nonumber \\
&& + \left( { K_\mathrm{eq} \over \Omega_0 } \right)^2 \left[
{\cal Z}_{1,1}+ {\cal Z}_{1,3}+ {\cal Z}_{2,2}
\right],
\label{eq:SI:Z}
\end{eqnarray}
where ${\cal Z}_{a}$ is the configurational free energy of a chain with two reacted complexes that are separated by $a\cdot L$ segments, while ${\cal Z}_{a,b}$ is the configurational free energy of a chain featuring two loops in which the reacted complexes are separated, before reacting, by $a\cdot L$ and $b\cdot L$ segments (see Tab.\ \ref{Tab:Table1}). In Eq.\ \ref{eq:SI:Z} the configurational partition function of unreacted ideal chains has been set at 1 (first line of Tab.\ \ref{Tab:Table1}). In particular, following Eq.\ \ref{Eq:Partfunc}, this implies that $\int_\varphi \mathrm{d} {\bf r}_2 \cdots \mathrm{d} {\bf r}_{N_C} =\prod_i \int_\varphi \mathrm{d} {\bf r}_i = 1\,\,\forall N_C$. The factors in front of ${\cal Z}_a$ are multiplicity terms counting the different ways of choosing two complexes distanced by $a\cdot L$ segments. \\
${\cal Z}_1$ can be calculated as follows 
\begin{eqnarray}
{\cal Z}_{1} &=& \int_\varphi \mathrm{d}{\bf r}_2 \cdots \mathrm{d} {\bf r}_{L} \delta(|{\bf r}_L-{\bf r}_1| -\sigma)
\label{eq:SI:Z1} \\ 
&=& \int \mathrm{d} {\bf r}_{L} p({\bf r}_{L} ,L) \delta(|{\bf r}_L-{\bf r}_1| -\sigma) = \Omega_0 p(0,L+1)
\nonumber
\end{eqnarray}
where we have used that $\int_\varphi \mathrm{d} {\bf r}_2 \cdots \mathrm{d} {\bf r}_{N_C} f({\bf r}_{N_C}) = \int \mathrm{d} {\bf r}_N p ({\bf r}_{N_C},L) f({\bf r}_{N_C})$ for any function $f$, along with the Markov property
\begin{eqnarray}
\int \mathrm{d} {\bf r}_L p({\bf r}_L,L) \delta (|{\bf r}_L - {\bf r}| -\sigma) 
&=& \int_{\Omega_0} \mathrm{d} {\bf u}\, p({\bf r}+{\bf u},L) 
\nonumber \\
&=& \Omega_0 \cdot p({\bf r},L+1) \, . 
\label{eq:SI:Markov}
\end{eqnarray}
In the previous equation the second integral is taken over the surface of the sphere centered in zero of radius $\sigma$ and area equal to $\Omega_0$.
Following the same steps leading to Eq.\ \ref{eq:SI:Z1} it can be shown that ${\cal Z}_{2} = \Omega_0 p(0, 2L+1)$, and that ${\cal Z}_{3} = \Omega_0 p(0,3L+1)$.

We now calculate the configurational partition functions of chains featuring two loops. ${\cal Z}_{1,1}$ (second line of Tab.\ \ref{Tab:Table1}) follows from the previous calculations as due to the fact that the configurational costs of forming two loops are independent: ${\cal Z}_{1,1} = [\Omega_0 p(0,L+1)]^2$. Similarly ${\cal Z}_{1,3}$ (third line of Tab.\ \ref{Tab:Table1}) can be calculated first reacting the smallest loop of length $L$, and then the biggest one: ${\cal Z}_{1,3}=\Omega_0^2\cdot p(0,L+1) \cdot p(0,2L+2)$. Note that after forming the first loop, the two complexes forming the biggest loop are separated by $2L+1$ segments. ${\cal Z}_{2,2}$ (fourth line of Tab.\ \ref{Tab:Table1}) is calculated, first, by reacting two complexes at distance $2L$ and then by reacting a dangle terminal of length $L$ with the middle point of the already formed loop 
\begin{eqnarray}
{\cal Z}_{2,2} 
&=& \Omega_0 p(0,2L+1)  \int \mathrm{d} {\bf r}_1   \mathrm{d} {\bf r}_2 \, p(|{\bf r}_1|,L) \cdot
\nonumber \\ && \qquad \qquad \qquad  \qquad
\cdot p^{(1)}_{L,L+1}(|{\bf r}_2|) \delta(|{\bf r}_1 - {\bf r}_2|-\sigma)
\nonumber \\
&=& \Omega_0 p(0,2L+1)  \int \mathrm{d} {\bf r}_2 \int_{\Omega_0} \mathrm{d} {\bf u} \, p(|{\bf r}_2+{\bf u}|,L) \cdot 
\nonumber \\ && \qquad \qquad \qquad  \qquad
\cdot p^{(1)}_{L,L+1}(|{\bf r}_2|) 
\nonumber \\
&=& \Omega_0^2 p(0,2L+1)  \int \mathrm{d} {\bf r}_2 \, p(|{\bf r}_2|,L+1) p^{(1)}_{L,L+1}(|{\bf r}_2|), \nonumber\\
\label{eq:S1:Z22}
\end{eqnarray}
where in the last equality we have used Eq.\ \ref{eq:SI:Markov} and $p^{(1)}_{L,L+1}$ is the distance probability of the $L$--ieme monomer of a loop made of $2L$ monomers tethered to the origin. In particular we have 
\begin{eqnarray}
p^{(1)}_{L,L+1}(|{\bf r}|) &=& {p(|{\bf r}|,L) p(|{\bf r}|,L+1) \over \int \mathrm{d}{\bf r}\, p(|{\bf r}|,L) p(|{\bf r}|,L+1) } \,\, .
\end{eqnarray}
Using the previous equation along with Eq.\ \ref{eq:S1:Z22} we obtain 
\begin{eqnarray}
{\cal Z}_{2,2} &=& \Omega_0^2  p(0,2L+1) { \int \mathrm{d}{\bf r} \, p(|{\bf r}|,L) p(|{\bf r}|,L+1)^2 \over \int \mathrm{d}{\bf r}\, p(|{\bf r}|,L) p(|{\bf r}|,L+1) } \,\, . \nonumber \\ 
\end{eqnarray}
The theoretical probabilities reported in Tab.~\ref{Tab:Table1} are given by $p_{0,0}=1/{\cal K}$, $p_{5,5}={\cal Z}_{1,1}/{\cal K}$, $p_{20,5}={\cal Z}_{1,3}/{\cal K}$, $p_{10,10}={\cal Z}_{2,2}/{\cal K}$ and using $L=5$ and ${\cal K}=1+{\cal Z}_{1,1}+{\cal Z}_{1,3}+{\cal Z}_{2,2}$.

\end{document}